# Impact Ejecta near the Impact Point Observed using Ultra-high-speed Imaging and SPH Simulations, and a Comparison of the Two Methods

**Takaya Okamoto[1†], Kosuke Kurosawa[2], Hidenori Genda[3], and Takafumi Matsui[2]**

[1] Institute of Space and Astronautical Science, Japan Aerospace Exploration Agency, Japan

[2] Planetary Exploration Research Center, Chiba Institute of Technology, Japan

[3] Earth–Life Science Institute, Tokyo Institute of Technology, Japan

Corresponding author: Takaya Okamoto (tokamoto@planeta.sci.isas.jaxa.jp)

† 3-1-1, Yoshinodai, Chuo-ku, Sagamihara, Kanagawa, 252-5210, Japan

**Key Points:**

- We performed 200 ns imaging of high-speed ejecta during oblique impacts.
- Computations under the same conditions as pertaining to the experiments reproduced the observed ejection behavior very well.
- Velocity boost owing to sustained compression works effectively for oblique impacts.

**Abstract**

    High-speed impact ejecta at velocities comparable to the impact velocity are expected to contribute to material transport between planetary bodies and deposition of ejecta far from the impact crater. We investigated the behavior of high-speed ejecta produced at angles of 45 and 90 degrees, using both experimental and numerical methods. The experimental system developed at the Planetary Exploration Research Center of Chiba Institute of Technology (Japan) allowed us to observe the initial growth of the ejecta. We succeeded in imaging high-speed ejecta at 0.2 µs intervals for impacts of polycarbonate projectiles of 4.8 mm diameter onto a polycarbonate plate at an impact velocity of ~4 km s$^{-1}$. Smoothed particle hydrodynamics (SPH) simulations of various numerical resolutions were conducted for the same impact conditions as pertaining to the experiments. We compared the morphology and velocities of the ejecta for the experiments and simulations, and we confirmed a close match for high-resolution simulations (with $\geq 10^6$ SPH particles representing the projectile). According to the ejecta velocity distributions obtained from our high-resolution simulations, the ejection velocities of the high-speed ejecta for oblique impacts are much greater than those for vertical impacts. The translational motion of penetrating projectiles parallel to the target surface in oblique impacts could cause long-term, sustained acceleration at the root of the ejecta.

**Plain-language Summary**

    Impact ejection is an inevitable outcome following hypervelocity impacts. Ejecta produced by the "spallation" process, which generates lightly shocked, high-speed materials, are thought to be important in the context of (Litho-)Panspermia, crater chronology, and the origin of Martian meteorites and tektites. The nature of impact spallation is not fully understood, especially in oblique impacts. In this study, we investigate the effects of the impact obliquity on spallation using both experimental and numerical approaches. The ejection behavior observed in the laboratory is well reproduced by our numerical code when we employ a sufficiently high spatial resolution. The experimentally validated numerical model allows us to address the nature of the impact-driven flow field. We found that the acceleration efficiency during oblique impacts is much better than during vertical impacts.

Index Term: 5420 Impact phenomena, cratering

**1. Introduction**

    Impact cratering is one of the major processes driving the geological evolution of the surfaces of planetary bodies in the solar system. Understanding the launch mechanism of materials from the vicinity of the impact point with high velocities is essential to address a number of long-standing problems in planetary science, including the origin of Martian meteorites (e.g., Head et al., 2002; Artemieva & Ivanov, 2004; Kurosawa et al., 2018), material exchange between planetary bodies (e.g. Chappaz et al., 2013; Ramsley & Head, 2013; Hyodo et al., 2019), the distribution of secondary craters (e.g. Plescia et al., 2010; Boyce & Mounginis-

Mark, 2015; Neesemann et al., 2016; Alvarellos et al., 2017; Xiao, 2018), and the origin of tektites (e.g., Koeberl, 1986; Vickery, 1993; Wasson, 2005).

Since oblique impacts are common (Shoemaker, 1963; Pierazzo & Melosh 2000), all these events are likely to occur for oblique impacts rather than for vertical impacts. To investigate whether materials from the parent bodies can be transferred to extremely distant regions and whether these materials can be deposited far away from their parent craters, it is first necessary to understand and obtain the ejecta velocity distributions of the high-speed ejecta during oblique impacts. The high-speed ejecta are launched from near the impact points, where the point-source approximation does not hold.

A number of laboratory experiments have been conducted to obtain the velocity distribution of impact ejecta from positions beyond the impactor's radius during vertical impacts (e.g., Cintala et al., 1999; Hermalyn & Schultz, 2011; Tsujido et al., 2015). The distribution of the ejection velocities as a function of the launch locus of the ejecta is a power-law relation, as expected from point-source scaling analysis (Housen & Holsapple, 2011). On the other hand, Anderson et al. (2003, 2004) conducted oblique impact experiments using three-dimensional particle image velocimetry (3D PIV), which allows direct measurements of the ejecta particle positions and their velocities. The ejection velocities and angles exhibit an asymmetric pattern with respect to the crater center, suggesting that the point source approximation is not valid to describe the subsurface flow field during oblique impacts. Ejected particles with velocities of up to ~3% of the impact velocity were measured, and the time variation of the velocity as a function of the azimuthal angle for an impact angle of 30° was shown in their studies. However, high-speed ejecta with velocities comparable to the impact velocities originating from just below the impact point for oblique impacts have not been studied in detail, which are important for transfer of materials to extremely far sides from impact points. Although the velocity of the highest part of the ejecta, observed as the leading edge of the ejecta plumes, has been measured from high-speed images (e.g., Kurosawa et al., 2015), it is difficult to obtain the velocity distribution pertaining to the subsequent continuous ejecta solely from laboratory experiments. On the other hand, numerical simulations would greatly help to address the entire ejection dynamics, sufficient comparisons of the behavior of high-speed ejecta between laboratory experiments and numerical simulations, however, have not been conducted.

Previous numerical studies clearly demonstrated that the behavior of materials moving at high velocities comparable to the impact velocity, including their masses and absolute particle velocities, are significantly affected by the spatial resolution employed in a simulation (Johnson et al., 2014; Kurosawa et al., 2018). Other numerical studies of high-speed ejecta after oblique impacts have shown that there are orders of magnitude more massive high-speed ejecta in oblique impacts than in vertical impacts (e.g., Shuvalov & Dypvik, 2004; Artemieva & Shuvalov, 2008; Shuvalov et al., 2012) and the masses of the high-speed ejecta are less sensitive to the spatial resolution in 20–100 cells per projectile radius (CPPR) (Artemieva & Ivanov, 2004). We should thus examine the spatial resolution required to reproduce the ejection dynamics of the materials originating near the impact point prior to the use of numerical codes.

High-speed ejecta are a product associated with the early stages of crater formation. Kurosawa et al. (2018) studied a spallation process for vertical impacts using numerical simulations. Ejecta launched by this spallation process are lightly shocked and accelerated to high velocities by "late-stage acceleration", which is triggered by the pressure gradient in the ejecta curtain. This acceleration mechanism is also expected to work at oblique impacts and thus may cause an increase in the ejection velocity. Note that we replace the term of "late-stage acceleration" used in Kurosawa et al. (2018) with "post-shock acceleration" in this study. The original term may be confusing because the term "late-stage" has been used to indicate the processes which are obvious at the end stage of the impact phenomena. In contrast, the "late-stage acceleration" introduced by Kurosawa et al. (2018) occurs at the beginning of the excavation stage. A brief explanation of the "post-shock acceleration" is described in Section 4.2.

In this study, we investigate high-speed ejecta using both experimental and numerical methods. We conducted laboratory impact experiments using an ultra-high-speed video camera with high temporal and spatial resolution. Next, numerical simulations using a smoothed particle hydrodynamics (SPH) code were performed at a variety of spatial resolutions under the same impact conditions as in the laboratory experiments. We explored the spatial resolution required to reproduce the behavior of the high-speed ejecta observed in the laboratory experiments through a comparison of the morphologies of the ejecta. Using the simulation data, validated by the laboratory experiments, we show the velocity distributions of the high-speed ejecta and discuss post-shock acceleration for oblique impacts compared with vertical impacts.

Note that in this paper we define materials ejected above the pre-impact target surface and with a velocity greater than ~500 m s$^{-1}$ as "ejecta", whereas previous numerical studies dealing with impact ejecta often used the term "ejecta" to indicate materials after a pressure release down to zero (or atmospheric) pressure.

The remainder of this manuscript is organized as follows. Section 2 describes the detail methods of the laboratory impact experiments and the numerical simulations. Section 3 presents their results, and a comparison of the laboratory experiments and the numerical simulations for different spatial resolutions. In Section 4, we describe the characteristics of the ejection behaviors of the high-speed ejecta. Our conclusions are summarized in Section 5.

## 2. Methods

### 2.1. Laboratory Impact Experiments

Impact experiments were conducted using a two-stage light-gas gun at the Planetary Exploration Research Center, Chiba Institute of Technology (PERC, Chitech), Japan (Kurosawa et al., 2015). We used polycarbonate plates (5×5×2 cm$^3$ or 5×10×2 cm$^3$) and a polycarbonate sphere with a diameter $d_p$ = 4.8 mm as target and projectile, respectively. We chose polycarbonate because of the following three reasons. First, the solidus temperature of polycarbonate (498 K; Haynes, 2010) is much lower than that of typical geological materials. Since the yield strength depends strongly on temperature and reduces to approximately zero at the melting point (Ohnaka, 1995), shocked polycarbonate near the impact point for our experimental condition is expected to behave as a perfect fluid (i.e., it is appropriate to simulate a high-speed impact of planetary material). Second, shock Hugoniot data for polycarbonate have been reported in the literature (Marsh, 1980). Finally, we can easily obtain a uniform medium made of polycarbonate. These three features allow us to obtain an idealized data set to compare with our numerical models.

We performed two oblique and one vertical impact experiments. The resulting impact velocities $v_i$ were 3.56 km s$^{-1}$ and 5.04 km s$^{-1}$ for the oblique impacts, and 4.18 km s$^{-1}$ for the vertical impact. Since the melting point of polycarbonate is much lower than that of rocky materials, as mentioned above, the impact velocities would correspond to somewhat higher impact velocities in natural impact events. To estimate the corresponding impact velocity for a collision between polycarbonate, we estimate the melt energy, that is, the specific internal energy on the intersection between the Hugoniot curve and the adiabat through the melting point at the reference pressure. The melt number $v_i^2/E_m$ is sometimes used to estimate the corresponding impact velocity. Figure 1 shows the melt number as a function of impact velocity for the laboratory impact events. We used $E_m$ = 8.7 MJ kg$^{-1}$ for basalt (Quintana et al., 2015) and $E_m$ = 2.5 MJ kg$^{-1}$ for polycarbonate. The $E_m$ for polycarbonate was calculated in the same manner used Quintana et al. (2015). The comparison shows that the impacts between polycarbonate in the laboratory at ~3.6 km s$^{-1}$ and ~7 km s$^{-1}$ correspond to natural impact events at ~6.7 km s$^{-1}$ and ~13 km s$^{-1}$, respectively, in terms of the melt number. Impact angles were set at 45 or 90 degrees measured from the target surface. An impact angle of 45 degrees corresponds to the most likely condition in natural impact events (Shoemaker, 1963; Pierazzo & Melosh 2000). The residual pressure in the vacuum chamber prior to each shot was <100 Pa. We placed a high-speed video camera (Shimadzu, HPV-X or HPV-X2) on one side of the chamber. A self-adjustable pre-event pulse generator (Kondo & Yasuo, 1987) was used to accurately adjust the camera's timing between the impact and observation. The experimental apparatus was described in detail by Kurosawa et al. (2015). The time interval between frames was set to 0.2 μs, which was shorter than a characteristic time for projectile penetration, $t_s = d_p/v_i$ (= 1.14 μs for the vertical impact at 4.18 km s$^{-1}$), to resolve the early stage of the material ejection. The experimental configuration is illustrated in Figure 2. The main difference from the previous study conducted at the same facility (Kurosawa et al., 2015) was the incidence direction of the light for the high-speed imaging. Although the previous study analyzed the motion of the self-luminous ejecta plumes, in this study we placed a light source (mecablitz 76 MZ-5, Digital flash or CAVITAR, Cavilux Smart System with a center wavelength of 640 nm) at the window on the opposite side to obtain backlight images during the impacts. When we used the Cavilux Smart System for the oblique impacts, a bandpass filter with a center wavelength of 640 nm was inserted in front of the camera lens, and the pulse duration of the light source was set to 50 ns. Self-illumination immediately after the impact is effectively suppressed using a single-wavelength filter and a light source of the same wavelength. Consequently, images of the high-speed ejecta during the initial ejection process could be captured clearly.

## 2.2. Numerical Simulations Using a Smoothed Particle Hydrodynamics Code

We used the SPH method (e.g., Lucy, 1977; Monaghan, 1992; Genda et al., 2015, 2017; Kurosawa et al., 2018), which is a flexible Lagrangian method used to solve hydrodynamic equations. The Tillotson (1962) equation of state (EOS) with the parameters for polycarbonate listed in Table 1 was used (Sugita & Schultz, 2003). A Von Neumann–Richtmyer-type artificial viscosity (Von Neumann & Richtmyer, 1950) with the standard parameter sets ($a_1 = 1.0$, $a_2 = 2.0$) was used to capture the shock waves in the numerical calculations. The material strength was neglected in the numerical model, because shocked polycarbonate near the impact point is expected to behave as a perfect fluid, as described in Section 2.1. Gravity was not included in the simulations either, because gravitational acceleration is expected to be negligible in the early stages of the ejection process.

The size of the projectile, the impact velocity, and the impact angle in the simulations were the same as those in the impact experiments. A spherical projectile was impacted onto the flat surface of a half-sphere target with a radius equivalent to the five-fold projectile radius. We can reduce the number of SPH particles for a half-sphere target, because the shock front travels near the impact point in a nearly spherical shape. The impact points were set in the center of the target for the vertical impacts, whereas they were set a projectile diameter away from the target's center for the oblique impacts. The target was sufficiently large to investigate the behavior of the ejecta launched by the end of the calculation. We chose the radius of the target hemisphere through trial and error, as follows. The effect of the expansion wave from the near-impact boundaries of the target hemisphere from the impact point—at the closest free surface from the impact point except for the top flat surface of the target—on the ejecta should be avoided to accurately investigate the ejection behavior. Given that the target hemisphere has a radius equivalent to the five-fold projectile radius, the time of shock arrival at the near-impact boundaries of the target hemisphere is close to the time defined as the travel distance between the impact point and the near-impact boundaries of the target hemisphere divided by the sound speed of polycarbonate under the conditions pertaining to our calculation. In this case, the time required for the expansion wave to travel back to the impact point was estimated as being twice the time for shock arrival at the boundaries (~4.8 μs). This period is equivalent to the end time of the calculation (4.8 μs). Consequently, the effects of the expansion wave, from the boundaries of the target hemisphere, on ejection behavior discussed in this study are negligible. Cartesian coordinates ($X, Y, Z$) were employed; $X$ for the direction towards the horizon, $Y$ for the horizontal direction perpendicular to the $X$ axis, and $Z$ for the height. In particular, for the oblique impacts, the $–X$ direction represented the direction of the horizontal component of the projectile trajectory. The impact points were taken as the origin of the coordinate system. SPH particles of the same mass were placed in a 3D lattice (face-centered cubic) within the sphere of a projectile and the half-sphere of the target (e.g., Fukuzaki et al., 2010). We set the number of SPH particles in a projectile, $n_{imp}$, to $10^4$, $10^5$, $10^6$, and $3\times10^6$ to investigate the effects of varying the spatial resolution on the ejection behavior. The CPPR corresponding to $n_{imp} = 10^4$, $10^5$, $10^6$, and $3 \times 10^6$ are approximately 13, 28, 62, and 89, respectively. The number of SPH particles in the target, $n_{target}$, increases with increasing $n_{imp}$. For example, $n_{target} \cong 6.3\times10^5$ and $1.9\times10^8$ at $n_{imp} = 10^4$ and $3\times10^6$, respectively. We adopted $t = 0$ for the time at the initial contact between the projectile and target, and we conducted simulations from $t = –0.2$ μs to $t = 4.8$ μs. The calculation conditions are summarized in Table 2.

## 2.3. Data Analysis

Data analyses of the simulation results were conducted using a particle tracking technique. We analyzed the stored data for each SPH particle, specifically the particle position, velocity, acceleration, density, pressure, and peak pressure experienced during the simulation.

To compare the morphologies of the ejecta in the laboratory experiments with those in the SPH simulations, we processed our images—both the high-speed images from the laboratory experiments and the data from the SPH simulations, whereas we also show some examples of pre-processed snapshots in Text S1 in the Supporting Information. Since the SPH particles are distributed in 3D, we produced a two-dimensional map of SPH particles projected onto the *X–Z* plane prior to image processing, using a procedure as described below. First, we selected the same spatial domain in the *X–Z* plane as that pertaining to the field of view of the camera and divided the space into 400×250 cells, identical to the number of pixels on the camera's detector. Next, the number of SPH particles included in each projected cell along the line of sight was counted. Figure 3 shows a schematic illustration of our data analysis.

Image processing was done using OpenCV (e.g., Bradski & Kaehler, 2008). We extracted the apparent edge of the ejecta in the images (see Section 3.2) to compare the results from the experiments with those from the simulations. We determined the location of the apparent edge using the technique of image binarization. We call the binary images pertaining to the laboratory experiments and numerical simulations 'EXP' and 'SPH' images, respectively. To produce the EXP images, background images for each shot (i.e., the images recorded immediately before the shot under the same filming conditions) were subtracted from the raw data. Following binarization, the outlines of the processed ejecta were extracted by employing additional image processing, referred to as the closing and opening. The details of our image processing are described in Text S2 in the Supporting Information.

## 3. Results

In this section, we mainly present our results for the oblique impact at 3.56 km s$^{-1}$, as well as the most important results for the other oblique impact (characterized by a higher impact velocity) and vertical impact. The detailed results of the latter two shots are presented in Texts S3 and S4 in the Supporting Information, respectively.

### 3.1. Results of the Laboratory Experiments and Numerical Simulations

Figure 4 shows high-speed video images of the ejected materials for the oblique impact. The asymmetry of the shape when comparing the downstream and upstream ejecta is obvious. The rear surface of the projectile was intact until at least $t = 0.8$ μs ($t = 0.59$ $t_s$). Two ejecta components are clearly observed from 1 μs ($t = 0.74$ $t_s$) following initial contact, as reported by previous studies (e.g., Schultz & Gault, 1990; Schultz et al., 2007). At this time, which is still less than the characteristic time ($t_s$), the spherical rear surface of the projectile disappeared. The first ejecta component corresponds to ricocheting material directed towards the downstream side of the projectile trajectory, with the velocity exceeding the impact velocity, and the other component expands above the impact point with a translational motion towards the downstream side. Hereafter, the former and latter ejecta components are referred to as "component 1" and "component 2", respectively. A kink structure was observed between the two components (hereafter referred to as the "kink"). We found that the transition from component 1 to component 2 occurs within 1.5 $t_s$ at 3.56 km s$^{-1}$. Figure 5 shows a schematic diagram of the two components and location of the kink.

Figure 6 shows images of the SPH simulations following the data analysis described in Section 2.3 for different $n_{imp}$ at $t = 3.0$ μs ($t = 2.2\ t_s$), along with an image from the impact experiment. The morphology of the ejecta in the simulation becomes more similar to that in the experiment for increasing $n_{imp}$. The two components and kink structure are reproduced for $n_{imp} \geq 10^6$. Images from the laboratory experiment and SPH simulations for the vertical impact are shown in Text S4.1 in the Supporting Information.

### 3.2. Comparisons of Laboratory Experiments and Numerical Simulations

We expected that the high-speed ejecta observed in this study would approximate a perfect fluid because of our use of the plastic projectile and target (see Section 2), resulting in the formation of a featureless, smooth shape of the ejecta. Thus, we did not identify distinct flying objects, such as the large grains observed in previous studies, in the images (e.g., Cintala et al., 1999; Tsujido et al., 2015). Therefore, we cannot measure their velocities. Instead, we measured the moving velocity of the locus of the apparent edge of the ejecta projected onto the X–Z plane. An example of this apparent edge of the ejecta is illustrated in panel (g) of Figure 4. We focus on the locus of the apparent edge in this study, because the ejected materials found on the edge are expected to have relatively high velocities. Image processing to obtain the apparent edges was briefly described in Section 2.3 and is fully discussed in Text S2 in the Supporting Information.

Figures 7 and 8 show the outlines of the ejecta after image processing for numerical simulations characterized by different $n_{imp}$ as well as the laboratory experiment for the oblique and vertical impacts, respectively. The morphologies of the apparent edge of the ejecta in the simulation for larger $n_{imp}$ is increasingly similar to that seen in the experiment. The position of the leading edge of the ejecta in Figure 7 was changed depending on $n_{imp}$. The loci in all simulations, however, are different from that observed in the experiment, indicating that the moving velocities of component 1 in the SPH simulations are slower than that observed in the laboratory experiment. An $n_{imp} > 3\times10^6$ would be necessary to reproduce the fastest part of the ejecta. On the other hand, for the vertical impact, the locus of the rising edge from the root of the ejecta in the experiment was well reproduced by all of our simulations.

To compare the results of the laboratory experiments and SPH simulations quantitatively, the moving velocities of the apparent edge of the ejecta were measured. Figure 9 shows the time variations of the loci of the apparent edge along the different survey lines. The slopes in this figure yield the relevant moving velocities. The simulation results for $n_{imp} = 3\times10^6$ are shown. Since the outlines of the edge were extracted from both the EXP and SPH images after image processing, adopting a thickness of ±1 pixel, there is an uncertainty associated with each data point. This was estimated as the length of the overlap between the survey line and outline of the ejecta. Figure 10 shows a comparison of the velocities of the moving edges from the laboratory experiment and SPH simulations. The slopes were obtained by employing the least-squares method. The uncertainty in the velocity comes from the uncertainties in the exact location and error in the fit. The velocities in the simulation for $n_{\mathrm{imp}} = 3\times10^6$ are in good agreement, to within ±15%, of the velocities in the laboratory experiment. Most parts of the apparent edge, except for the leading edge (as mentioned above), are reproduced well in the SPH simulation. This implies that the spatial and velocity distributions of these parts in the experiment are very close to their counterparts in the simulation. Note that the difference in the moving velocities in the laboratory experiment due to different values for the binarization threshold in the image processing is within 4% (see Text S2 in the Supporting Information), which is smaller than the 15% difference between the velocities in the experiments and numerical results, showing that the edge velocities do not depend significantly on the binarization threshold adopted in the image processing. The velocities in the simulation for $n_{\mathrm{imp}} = 10^6$ are also in agreement, to within ±30%, of the velocities in the laboratory experiment, and in particular those at angles <30 degrees are in agreement to within ±15%, just like for $n_{\mathrm{imp}} = 3\times10^6$, suggesting that the apparent edge around the kink, on which we focus in particular in this study, could be reproduced adequately at this resolution.

The moving velocities of the apparent edges for the oblique impact at the higher impact velocity of 5.04 km s$^{-1}$ and the vertical impact at an impact velocity of 4.18 km s$^{-1}$ were also measured and are shown in Figures S7 and S11 in the Supporting Information, respectively. In both cases, most parts of the edges in the laboratory experiments are also reproduced well by those in the SPH simulations of high spatial resolution. In the former case, the moving velocities for $n_{\mathrm{imp}} \geq 10^6$ are in good agreement, to within approximately ±15%, of the velocities in the laboratory experiment, except for the impacts at angles <30 degrees. In the vertical impact case, the velocities at impact angles <75 degrees in our simulations for $n_{\mathrm{imp}} \geq 10^6$ are reproduced to within ±30% of the velocities in the laboratory experiment. Especially for $n_{\mathrm{imp}} = 3\times10^6$, the velocities in the simulation reproduced the velocities in the laboratory experiment to within ±15%. Therefore, we show that the moving velocities of the edge are in good agreement to within an accuracy of ±15% between the results from the laboratory experiments and the SPH simulations for $n_{\mathrm{imp}} = 3\times10^6$ at both the higher and lower impact velocities and for typical impact angles of 45 and 90 degrees. Thus, the spatial and velocity distributions of most of the high-speed ejecta in the simulation of high spatial resolution, $n_{\mathrm{imp}} = 3\times10^6$, are very close to those in the experiments, if hypervelocity impacts occur with certain impact angles.

## 4. Discussion

In the previous section, we showed that our numerical simulations reproduced the spatial distribution of the high-speed ejecta and moving velocity of the edges of the ejecta for $n_{\mathrm{imp}} \geq 10^6$. This implies that it is necessary for calculations with >~60 CPPR (= $n_{\mathrm{imp}} \geq 10^6$) to reproduce the spatial and velocity distributions of the high-speed ejecta in the laboratory experimental results, whereas it may be sufficient for the masses of the high-speed ejecta to be calculated with a resolution of 20 CPPR, because Artemieva & Ivanov (2004) suggested that the resulting masses in calculations with 20 and 100 CPPR, using a different code (SOVA), are almost the same.

There is a great advantage in the use of numerical simulations, in that, in-depth information, such as the 3D velocity distribution and time variation of a pressure field can be obtained. We reconstruct the 3D velocity distributions in the laboratory impact experiment using the 3D SPH code in Section 4.1 and discuss possible production processes of the high-speed ejecta observed in the experiments in Section 4.2 and how effectively post-shock acceleration worked for oblique impacts in Section 4.3.

## 4.1. The Velocity Distribution

In order to understand the transfer of material between planetary bodies, as well as the distribution and thickness of ejecta deposits far from the impact point, it is necessary to obtain initial launch conditions. Our numerical model validated by the experimental results could provide such initial 3D velocity distributions of ejecta and their features. In Section 4.1, the 3D velocity distributions and initial launch positions in the laboratory experiments were investigated using the results of the SPH simulations for $n_{\mathrm{imp}} = 3 \times 10^6$.

We extracted the SPH particles from the upper edge of the ejecta, which are expected to have relatively large velocities, to investigate the velocity distributions. Such detailed analyses are difficult to undertake successfully based on laboratory experiments. Figure 11 shows the extracted SPH particles at $t = 3.0$ μs ($t = 2.2\ t_{\mathrm{s}}$) projected onto the $X$–$Z$ and $X$–$Y$ planes. The SPH particles are color-coded depending on the ejection velocity $v_{\mathrm{ej}}$, normalized by the impact velocity. The ejecta from the target and projectile materials are shown in panels (a) and (b), and in panels (c) and (d) of Figure 11, respectively. We found that most of the ejected particles forming the apparent edge have velocities greater than the impact velocity. Here, the extracted SPH particles are divided into three groups, "component 1" ($X \leq -20$ mm), the "kink" ($-15$ mm $< X \leq -20$ mm), and "component 2" ($-15$ mm $< X$). Component 1 corresponds to the fastest fraction of the ejecta. The SPH particles in the kink have moderate ejection velocities and form the unique kink structure. Component 2 corresponds to the distorted hemispherical component mentioned in Section 3.1. Although our SPH model does not reproduce the fastest part of the ejecta (see Section 3.2), the leading edge seems to come predominantly from the projectile. The kink is also produced mainly by the projectile-derived ejecta. In component 2, the projectile materials are concentrated on the far side of the impact point in the horizontal direction of the projectile trajectory, and the target materials are concentrated on the near side of the impact point. The cumulative mass of the high-speed ejecta as a function of ejection velocity $v_{\mathrm{ej}}$, is shown in Figure S15 in the Supporting Information.

Figure 12 shows the distribution of the direction vectors of the ejection velocities. The directions are characterized by the elevation angle, $\theta$, and the azimuth angle, $\phi$. The elevation angle is defined as the angle measured from the target surface. The azimuth angle is defined as the angle measured from the projectile trajectory in a clockwise fashion. For example, $\phi = 0°$ corresponds to the $-X$ direction and $\phi = 90°$ corresponds to the $+Y$ direction. The relevant ranges of $\theta$ and $\phi$ are $0° \leq \theta \leq 90°$ and $-180° < \phi \leq 180°$, respectively. Examples for a number of combinations of ejection angles, $(\theta, \phi) = (30, 45), (30, -45), (30, 135), (30, -135)$ [degrees] are illustrated in Figure 12(a). The range $-90° < \phi < 90°$ represents the direction of the ejecta moving to the downstream side of the impact point, whereas the ranges of $-180° < \phi < -90°$ and $90° < \phi < 180°$ represent the direction of the ejecta moving to the upstream side of the impact point. Note that only the range $0° \leq \phi \leq 180°$ is shown in Figure 12(b)–(c), because the distribution exhibits mostly an axially symmetric pattern with respect to the $X$ axis.

We characterize the ejecta distribution in this diagram as follows. A large fraction of the target materials is distributed across 30° ≤ $\phi$ ≤ 150°. The ejection velocity decreases with increasing $\phi$. On the other hand, the projectile materials are distributed approximately at 0° ≤ $\phi$ ≤ 90°, suggesting that almost all projectile-derived ejecta move to the downstream side of the projectile trajectory. Target materials dominate at 60° ≤ $\phi$ ≤ 90° on the downstream side compared with projectile materials. Particles moving at ejection velocities twice the impact velocity are distributed at $\theta$ < 20° and $\phi$ < 60°. Our 3D initial velocity distribution also shows that there are high-speed ejecta with extremely large ejection angles ($\theta \cong 90°$) for oblique impacts.

Initial positions of the extracted SPH particles from the target are shown in Figure 13. The initial positions of the ejecta in component 1 are located in the top layer of the target and on the downstream side of the projectile trajectory. The positions of the ejecta in the kink are also located only in the first and second layers of the target. The particles in the kink were ejected from a position closer to the impact point and slightly removed from the $X$ axis compared with those in component 1. On the other hand, the particles in component 2 were ejected from the target's first to fifth layers. They are located at positions at about one projectile radius from the $X$ axis. A fraction of the particles originating from the projectile in component 2 was ejected from the upstream side of the projectile trajectory. The results for the vertical impact corresponding to Figures 11, 12, and 13 are shown in Text S4.2 in the Supporting Information. These results also show that the extracted particles from the top apparent edge of the ejecta were ejected from the vicinity of the impact point (i.e., the target's first to sixth layers). Kurosawa et al. (2018) showed that the peak pressures of materials initially located in the top three layers below the target surface in the 3D SPH model are systematically lower than the expected values, because they are subjected to a shock smearing owing to the artificial viscosity (e.g., De Carli, 2013). Nevertheless, we found that the distributions of the positions and ejection velocities of the particles were similar to those in the laboratory experiments, indicating that our SPH model can reproduce the hydrodynamic behavior of the particles initially located even near the surface.

## 4.2. Peak Pressure vs Ejection Velocity

The peak pressure pertaining to each particle was calculated in the simulation. Figure 14 shows the ejection velocities of the extracted particles at the top edges of the ejecta for the oblique and vertical impacts as a function of the peak pressure. For the oblique impact, many particles in component 1, the kink, and component 2 are ejected at velocities greater than the impact velocity. These particles are defined as jets, based on previous studies where jets squirt from the contact point of the target and projectile, reaching ejection velocities $v_{ej} > v_i$ (e.g., Johnson et al., 2014). Note that we can rule out the possibility that the high ejection velocities observed in the simulation are artifacts caused by our simplified treatment of vaporization in the Tillotson EOS (see Text S6 in the Supporting Information). Although no jets were found in the extracted particles on the apparent edge for the vertical impact, trace amounts of particles located on the outside of the apparent top edge have velocities greater than the impact velocity. If SPH simulations at a higher spatial resolution are performed to simulate the vertical impact, it is expected that jets would be also found on the apparent edge.

Here, we briefly discuss the expected upper limit to the ejection velocity as a function of the particle velocity immediately after shock arrival, $u_{pH}$, based on shock physics. The directions of material velocity vectors and their absolute particle velocities in an excavation flow can be estimated as the sum of the velocity vectors driven by a shock wave and an expansion wave. The theoretical maximum particle velocities in the case of the presence of a shock wave can reach twice the particle velocity of the shocked state, $u_{pH}$ (e.g., DeCarli, 2013). This is known as the "velocity-doubling rule" at the free surface. Based on a geometric consideration of the impact-driven flow field for vertical impacts, the maximum ejection velocity would be reduced to $\sqrt{2}u_{pH}$ (Kurosawa et al., 2018). Nevertheless, recent high-resolution model calculations performed by Kurosawa et al. (2018) showed that the ejection velocity could be $> 2u_{pH}$ because of what they referred to as "late-stage acceleration". As mentioned in Section 1, here we refer to this process as "post-shock acceleration". Post-shock acceleration is more gradual acceleration by the compressive nature of the ejection flow near the impact point. Such gradual acceleration is driven by a pressure gradient produced by sustained compression around the target surface. The pressure gradient is directed upward and outward, leading to efficient acceleration. This is a unique feature of post-shock acceleration. The sustained compression occurs inevitably, because there is the large difference in the particle velocities near the impact point owing to the decaying shock wave (Kurosawa et al., 2018). Note that "sustained compression" was originally referred to as "material pile-up" by Kurosawa et al. (2018). Figure 14 implies that the acceleration efficiency in oblique impacts is much higher than in vertical impacts. In the next section, we consider the effect of the impact's obliquity on the degree of acceleration.

## 4.3. Post-Shock Acceleration in Oblique Impacts

In this section, we discuss the fundamental differences between oblique and vertical impacts and their effects on the efficiency of post-shock acceleration. In oblique impacts, a projectile has some translational velocity parallel to the target surface, resulting in a lower degree of shock compression. This means that the energy and momentum transfer from the projectile to the target in oblique impacts through shockwave passage is rather low compared with that associated with vertical impacts for the same impact velocities. The penetrating projectile, however, still has a translational velocity parallel to the target surface, causing the target materials to extrude in front of the projectile. Thus, the near-surface excavation flow in front of the projectile will prevent materials from decompressing adiabatically, prolonging the period of the elevated pressure. If the remaining translational velocity of the projectile is greater than the particle velocities of the materials in front of the projectile, the translational motion enhances the degree of the sustained compression. To investigate the strength and duration of sustained and of post-shock acceleration in oblique impacts, we analyzed the ejection behavior of the selected SPH particles in detail. Figure 15 shows the time variations of the particle velocities, accelerations, and temporal pressures of the selected particles with velocities $> 2u_{pH}$ in the last time step (4.8 μs). These particles are initially located on the near-target surface, within 15 SPH layers from the free surface and at the cross-section of the $X$–$Z$ plane for $Y < \pm 0.022$ mm, closest to the $Y = 0$ plane in the simulation. The initial spikes in the acceleration correspond to shock arrivals. Subsequently, they exhibit pressure plateaus during pressure release rather than a monotonic decrease in the temporal pressures, as also observed by Kurosawa et al. (2018). The materials gradually accelerated to $>1$ km s$^{-1}$ over periods of 0.5–1.0 $t_s$, even when the particle velocities immediately after shock arrivals are < 0.4 km s$^{-1}$. Figure 16 shows snapshots of close-up views around the edges of the projectile footprints. Colors indicate temporal pressure (not peak pressure) of the particles at each time step for oblique and vertical impacts (only for particles at the cross-section of the *X–Z* plane for *Y* < ±0.022 mm). We found that the pressure at the root of the ejecta for the oblique impact was still > ~1 GPa at *t* = 1.9 $t_s$. In contrast, the pressure pertaining to the vertical impact above the target surface is much lower, even at *t* = 0.52 $t_s$. Figures 15 and 16 suggest that post-shock acceleration is expected to be more significant for oblique than for vertical impacts because of the long duration of the acceleration. This long duration is expected to ultimately originate from the translational motion of the projectile parallel to the target surface in oblique impacts.

A first-order estimation of the velocity boost owing to post-shock acceleration, Δ*v*, was given by Kurosawa et al. (2018):

$$\Delta v = a_{\text{late}} \Delta t = \frac{1}{\rho} \frac{\partial P}{\partial r} \Delta t = \frac{P_{\text{root}}}{\rho L} \Delta t, \quad (1)$$

where $a_{\text{late}}$, Δ*t*, $\rho \cong 1200$ kg m$^{-3}$, $P_{\text{root}}$, and *L* are the magnitude and duration of the acceleration, density of the target materials in the ejection flow, pressure at the root of the ejecta curtain, and thickness of the ejecta curtain, respectively. Note that we assumed $\rho$ to be approximated by the reference density. Using typical results obtained from the numerical calculations, Δ*v* for the oblique impact at an impact velocity of 3.56 km s$^{-1}$ is expressed as:

$$\Delta v = 4.2 \left(\frac{P_{\text{root}}}{1\,[\text{GPa}]}\right)\left(\frac{\Delta t}{1.9\,[t_s]}\right)\left(\frac{\rho}{1200\,[\text{kg m}^{-3}]}\right)^{-1}\left(\frac{L}{0.21\,[r_p]}\right)^{-1} [\text{km s}^{-1}]. \quad (2)$$

We also estimated Δ*v* for the vertical impact and an impact velocity of 4.18 km s$^{-1}$, i.e.,

$$\Delta v = 1.7 \left(\frac{P_{\text{root}}}{1\,[\text{GPa}]}\right)\left(\frac{\Delta t}{0.35\,[t_s]}\right)\left(\frac{\rho}{1200\,[\text{kg m}^{-3}]}\right)^{-1}\left(\frac{L}{0.083\,r_p}\right)^{-1} [\text{km s}^{-1}]. \quad (3)$$

Our simple estimation suggests that the velocity boost owing to post-shock acceleration would be sufficiently large to explain the excess of the ejection velocity with respect to the $2u_{\text{pH}}$ lines shown in Figure 14.

## 5. Conclusions

We performed both laboratory experiments and numerical simulations to investigate the production processes of high-speed ejecta and their velocity distribution. Here, we considered the effects of impact obliquity. A high-speed video camera with a frame rate shorter than the characteristic time of projectile penetration was used to observe the initial growth of impact ejecta in oblique impacts. The two-dimensional projected shape obtained from the laboratory experiments can be reproduced by the 3D SPH code for $n_{\text{imp}} \geq 10^6$. In particular, the velocities at the edges of the ejecta in the simulations for different elevation angles at $n_{\text{imp}} = 3 \times 10^6$ are in good agreement, to within ±~15%, with the velocities measured in the laboratory experiments for both the oblique and vertical impacts. We demonstrated the reconstruction of the 3D velocity distributions in the laboratory impact experiment using the 3D SPH code with a reliable resolution of $n_{\mathrm{imp}} = 3 \times 10^6$. It showed that the ejection velocities at the apparent edge of the ejecta in the oblique impact were much higher than those for the vertical impact at a comparable impact velocity. The translational motion of the penetrating projectile parallel to the target surface could produce a strong pressure gradient in the materials in front of the projectile, which locates at the root of the ejecta, leading to a much higher ejection velocity than for the vertical impacts at the same impact velocity.

**Acknowledgments and Data**

We appreciate useful discussions during workshops on planetary impacts held at the Institute of Low Temperature Science, Hokkaido University, Japan, and at the Department of Planetology, Kobe University, Japan. We thank Hiroki Senshu (PERC, Chitech) for his support and allowing us to use a computer equipped at PERC for post-data analyses of the SPH calculations. We would also like to express our gratitude to the reviewer for the critical reviews which have improved the manuscript greatly and Gareth Collins for the constructive comments as an Associate Editor. This research was supported by Japan Society for the Promotion of Science (JSPS) Kakenhi Grant No. JP17H02990. TO is supported in part as JSPS Research Fellow, No. JP18J00027. KK is supported by JSPS Kakenhi Grant Nos. JP17K18812, JP17H01176, and JP17H01175. HG is supported by MEXT Kakenhi Grant No. JP17H06457. Data used in Figures from the results of ultra-high-speed imaging and SPH simulations are available by accessing the data repository site of Chiba Institute of Technology, http://id.nii.ac.jp/1196/00000228/.

**References**

Alvarellos, J. L., Dobrovolskis, A. R., Zahnle, K. J., Hamill, P., Dones, L., & Robbins, S. (2017). Fates of satellite ejecta in the Saturn system, II. *Icarus*, 284, 70–89.

Anderson, J. L. B., Schultz, P. H., & Heineck, J. T. (2003). Asymmetry of ejecta flow during oblique impacts using three-dimensional particle image velocimetry. *J. Geophys. Res. Planets*, 108, E8. doi: 10.1029/2003JE002075.

Anderson, J. L. B., Schultz, P. H., & Heineck, J. T. (2004). Experimental ejection angles for oblique impacts: implications for the subsurface flow-field. *Meteorit. Planet. Sci.*, 39, 303–320.

Artemieva, N., & Ivanov, B. (2004). Launch of Martian meteorites in oblique impacts. *Icarus*, 171, 84–101.

Artemieva, N. A., & Shuvalov, V. V. (2008). Numerical simulation of high-velocity impact ejecta following falls of comets and asteroids onto the Moon. *Solar System Research*, Vol. 42, No. 4, 329-334.

Bradski, G., & Kaehler, A. (2008). *Learning OpenCV: Computer vision with the OpenCV library*. Sebastopol, CA: O'Reilly Media, Inc.

Boyce, J. M., & Mouginis-Mark, P. J. (2015). Anomalous areas of high crater density on the rim of the Martian crater Tooting. *Abstracts of the workshop on issues in crater studies and the dating of planetary surfaces*, LPI Contribution No. 1841, p. 9005.

Chappaz, L., Melosh, H. J., Vaquero, M., & Howell, K. C. (2013). Transfer of impact ejecta material from the surface of Mars to Phobos and Deimos. *Astrobiol.*, 13, 963–980.

Cintala, M. J., Berthoud, L., & Hörz, F. (1999). Ejection-velocity distributions from impacts into coarse-grained sand. *Meteorit. Planet. Sci.*, 34, 605–623.

De Carli, P. S. (2013). Meteorites from Mars via a natural two-stage gas gun. *Proc. Eng.*, 58, 570–573.

Fukuzaki, S., Sekine, Y., Genda, H., Sugita, S., Kadono, T., & Matsui, T. (2010). Impact-induced $N_2$ production from ammonium sulfate: Implications for the origin and evolution of $N_2$ in Titan's atmosphere. *Icarus*, 209, 715–722.

Genda, H., Kobayashi, H., & Kokubo, E. (2015). Warm debris disks produced by giant impacts during terrestrial planet formation. *Astrophys. J.*, 810, 136.

Genda, H., Fujita, T., Kobayashi, H., Tanaka, H., Suetsugu R., & Abe Y. (2017). Impact erosion model for gravity-dominated planetesimals. *Icarus*, 294, 234–246.

Haynes, W. M. (2010). *CRC Handbook of Chemistry and Physics*, 91st ed., Boca Raton, FL: CRC Press.

Head, J. N., Melosh, H. J., & Ivanov, B. A. (2002). Martian Meteorite Launch: High-Speed Ejecta from Small Craters. *Science*, 298, 1752–1756.

Housen, K. R., & Holsapple, K. A. (2011). Ejecta from impact craters. *Icarus*, 211, 856–875. doi:10.1016/j.icarus.2010.09.017.

Hyodo, R., Kurosawa, K., Genda, H., Usui, T., & Fujita, K. (2019). Transport of impact ejecta from Mars to its moons as a means to reveal Martian history. *Sci Rep* 9, p. 19833.

Johnson, B. C., Bowling, T. J., & Melosh, H. J. (2014). Jetting during vertical impacts of spherical projectiles. *Icarus*, 238, 13–22.

Koeberl, C. (1986). Geochemistry of tektites and impact glasses. *Annu. Rev. Earth Planet. Sci.*, 14, 323–350.

Kondo, K., & Yasuo, H. (1987). Self-adjustable pre-event pulse generator for shock wave experiments. *Rev. Sci. Instrum.*, 58, 1755–1757.

Kurosawa, K., Nagaoka, Y., Senshu, H., Wada, K., Hasegawa, S., Sugita, S., & Matsui, T. (2015). Dynamics of hypervelocity jetting during oblique impacts of spherical projectiles investigated via ultrafast imaging. *J. Geophys. Res. Planets*, 120, 1237–1251.

Kurosawa, K., Okamoto, T., & Genda, H. (2018). Hydrocode modeling of the spallation process during hypervelocity impacts: Implications for the ejection of Martian meteorites. *Icarus*, 301, 219–234.

Lucy, L. B. (1977). A numerical approach to the testing of the fission hypothesis. *Astrophys. J.*, 82, 1013–1024.

Marsh, S. P. (1980). *LASL Shock Hugoniot Data* (658 pp.). Berkeley, CA: Univ. of Calif. Press.

Melosh, H. J. (1989). *Impact Cratering: A Geologic Process*. New York, NY: Oxford Univ. Press.

Monaghan, J. J. (1992). Smoothed particle hydrodynamics. *Annu. Rev. Astron. Astrophys.*, 30, 543–574.

Neesemann, A., Kneissl, T., Schmedemann, T., Walter, S. H. G., Michael, G. G., van Gasselt, S., Hiesinger, H., Jaumann, R., Raymond, C., & Russell, C. T. (2016). Size-frequency distributions of km to sub-km sized impact craters on Ceres. *Abstracts of 47th lunar and planetary science conf.*, LPI Contribution No. 1903, p. 2936.

Ohnaka, M. (1995). A shear failure strength law of rock in the brittle-plastic transition regime. *Geophys. Res. Lett.*, 22, 25–28.

Pierazzo, E., Vickery, A. M., & Melosh, H. J. (1997). A Reevaluation of Impact Melt Production. *Icarus*, 127, 408–423.

Plescia, J. B., Robinson, M. S., & Paige, D. A. (2010). Giordano Bruno: the young and the restless. *Abstracts of 41st lunar and planetary science conf.*, LPI Contribution No. 1533, p. 2038.

Quintana, S. N., Crawford, D. A., & Schultz, P.H. (2015), Analysis of Impact Melt and Vapor Production in CTH for Planetary Applications , *Procedia Engineering* 103, 499–506.

Ramsley, K. R., & Head, J. W. (2013). Mars impact ejecta in the regolith of Phobos: bulk concentration and distribution. *Planet. Space Sci.*, 87, 115–129.

Schnell, H. (1964), *Chemistry and Physics of Polycarbonates*, John Wiley, New York, 225 pp.

Schultz, P. H., Eberhardy, C. A., Ernst, C. M., A'Hearn, M. F., Sunshine, J. M., & Lisse, C. M. (2007). The deep impact oblique impact cratering experiment. *Icarus*, 190, 295–333.

Schultz, P. H., & Gault, D. E. (1990). Prolonged global catastrophes from oblique impacts. *Geol. Soc. Am. Spec. Pap.*, 247, 239–261.

Shoemaker, E. M. (1963). Impact mechanics at Meteor Crater, Arizona. In B. M. Middlehurst, G. P. Kuiper (Eds.), *The Solar System* (Vol. 4, pp. 301–336). Chicago, IL: Univ. of Chicago Press.

Sugita, S., & Schultz, P. H. (2003). Interactions between impact-induced vapor clouds and the ambient atmosphere: 2. Theoretical modeling. *J. Geophys. Res.*, 108, 5052.

Tillotson, J. H. (1962). Metallic equations of state for hypervelocity impact. *Tech. Rep. GA-3216* (General Atomic Report).

Thompson, S. L., & Lauson, H. S. (1972). Improvements in the CHART D radiation-hydrodynamic code III: Revised analytic equations of state. *Sandia National Laboratory Report* SC-RR-71 0. Albuquerque, New Mexico: Sandia National Laboratory. 113 p.

Tsujido, S., Arakawa, M., Sizuki, A. I., & Yasui, M. (2015). Ejecta velocity distribution of impact craters formed on quartz sand: effect of projectile density on crater scaling law. *Icarus*, 262, 79–92.

Vickery, A. M. (1993). The theory of jetting: application to the origin of tektites. *Icarus*, 105, 441–453.

Von Neumann, J., & Richtmyer, R. D. (1950). A method for the numerical calculation of hydrodynamic shocks. *J. Appl. Phys.*, 21, 232–237.

Wasson, J. T. (2015). *An origin of splash-form tektites in impact plumes*. Paper presented at 46th Lunar and Planetary Science Confer., Houston, TX. Abstract #1832 (CD-ROM).

Xiao, Z.-Y. (2018). On the importance of self-secondaries. *Geosci. Lett.*, 5–17.

**Table 1.** Input parameters for the Tillotson EOS for polycarbonate[a] (Sugita & Schultz, 2003).

| Parameter | Value |
| --- | --- |
| Reference density (kg m$^{-3}$) | 1194 |
| Tillotson constant, $a$ | 0.5 |
| Tillotson constant, $b$ | 1.0 |
| Bulk modulus, $A$ (GPa) | 9.2 |
| Tillotson constant, $B$ (GPa) | 6.9 |
| Tillotson constant, $E_0$ (MJ kg$^{-1}$) | 2 |
| Tillotson constant, $\alpha$ | 5 |
| Tillotson constant, $\beta$ | 5 |
| Specific internal energy for incipient vaporization, $E_{iv}$ (MJ kg$^{-1}$) | 0.28 |
| Specific internal energy for complete vaporization, $E_{cv}$ (MJ kg$^{-1}$) | 1.3 |

[a] Notations for the Tillotson parameters are the same as in Tillotson (1962).

**Table 2.** Setup parameters for the SPH calculations for polycarbonate.

|  | Oblique impacts | | Vertical impact |
| --- | --- | --- | --- |
| Impact angle (degrees) | 45 | | 90 |
| Impact velocity, $v_i$ (km s$^{-1}$) | 3.56 | 5.04 | 4.18 |
| Projectile radius, $r_p$ (mm) | 2.4 | 2.4 | 2.4 |
| Target radius, $r_t$ (mm) | 12.0 | 12.0 | 12.0 |
| Number of SPH particles[a] | $10^4$ (640,490), $10^5$ (6,375,728), $10^6$ (63,619,967), $3\times10^6$ (190,749,263) | $10^5$ (6,375,728), $10^6$ (63,619,967), $3\times10^6$ (190,749,263) | $10^4$ (640,490), $10^5$ (6,375,728), $10^6$ (63,619,967), $3\times10^6$ (190,749,263) |

[a] Number of SPH particles in the projectile. The corresponding CPPR to the number of SPH particles in a projectile of $10^4$, $10^5$, $10^6$, and $3\times10^6$ are approximately 13, 28 , 62, and 89, respectively. The total number of SPH particles is given in parentheses.

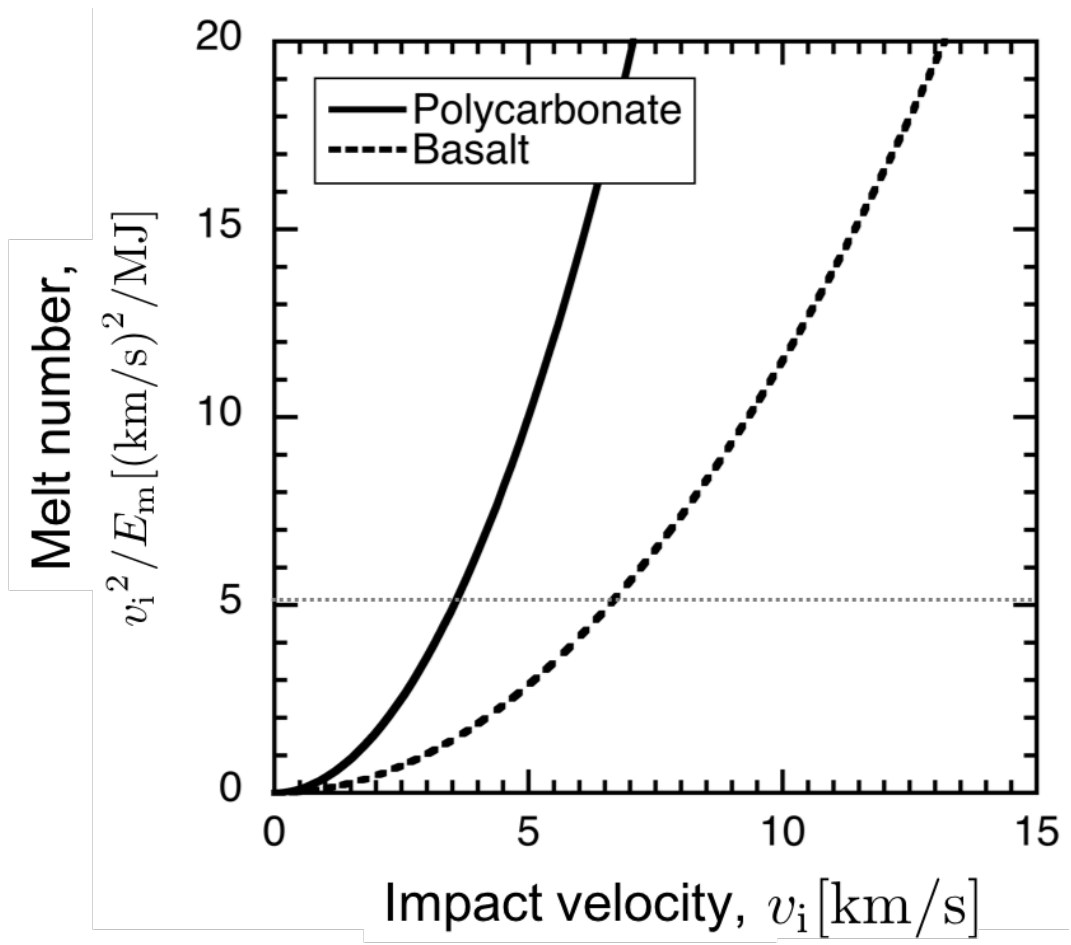

**Figure 1.** Melt number as a function of impact velocity. Solid and dashed lines are melt number for polycarbonate and basalt, respectively. The horizontal dotted line is the line of melt number of ~5, which corresponds to the impact velocity of ~3.6 km s$^{-1}$ for the impact experiments in this study, and ~6.7 km s$^{-1}$ for impact events on the Earth.

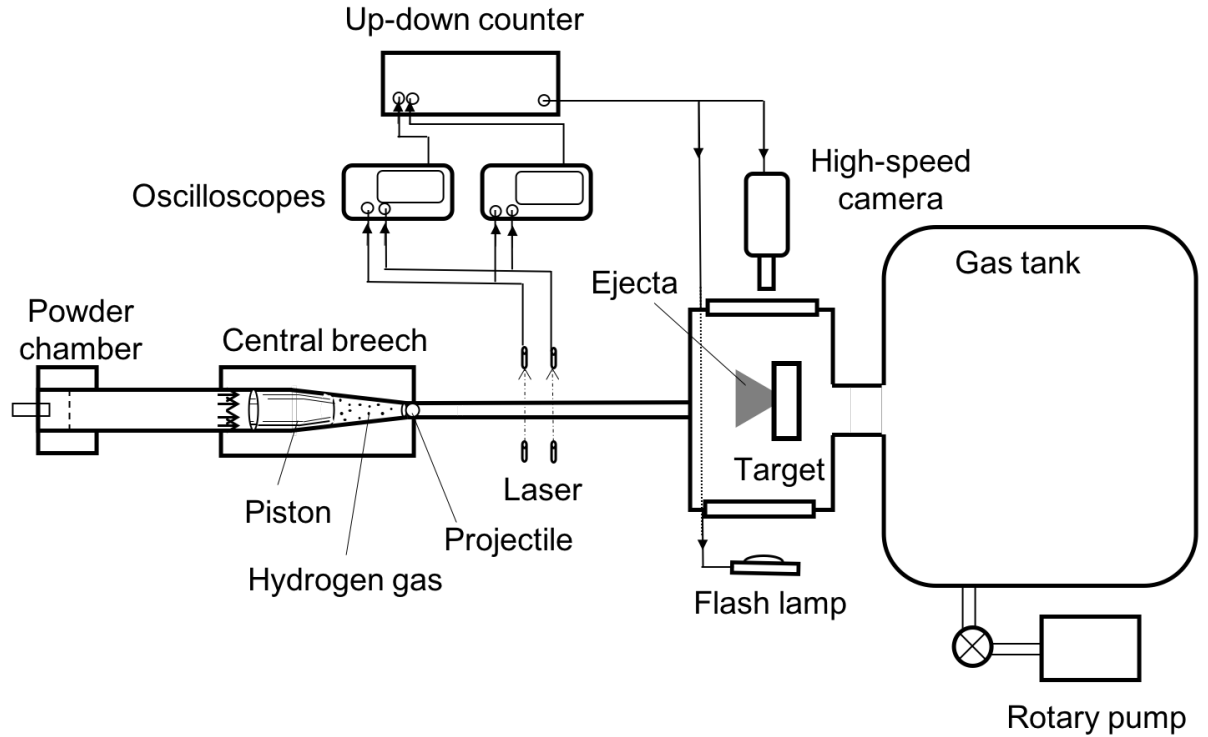

**Figure 2.** Schematic diagram of the experimental setup.

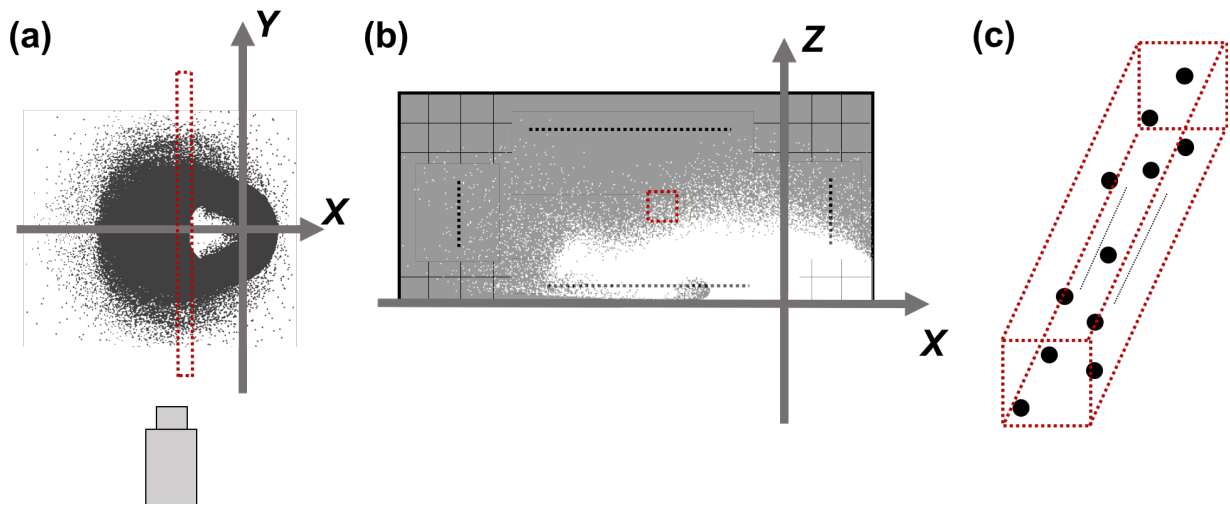

**Figure 3.** Schematic diagram of our data analysis of the SPH simulations. We only show the range $Z > 0$. (a) View of the ejected particles from the $+Z$ direction. The dotted red rectangle shows an example of an enlarged space of a single pixel in a high-speed image extending along the line-of-sight direction from the camera. (b) View of the ejected particles from the $-Y$ direction. The red square shows an example of a single pixel space in a high-speed image. (c) 3D enlarged view of the dotted red rectangles shown in panels (a) and (b). The black dots show examples of SPH particles in this space. SPH particles included in each cube were counted.

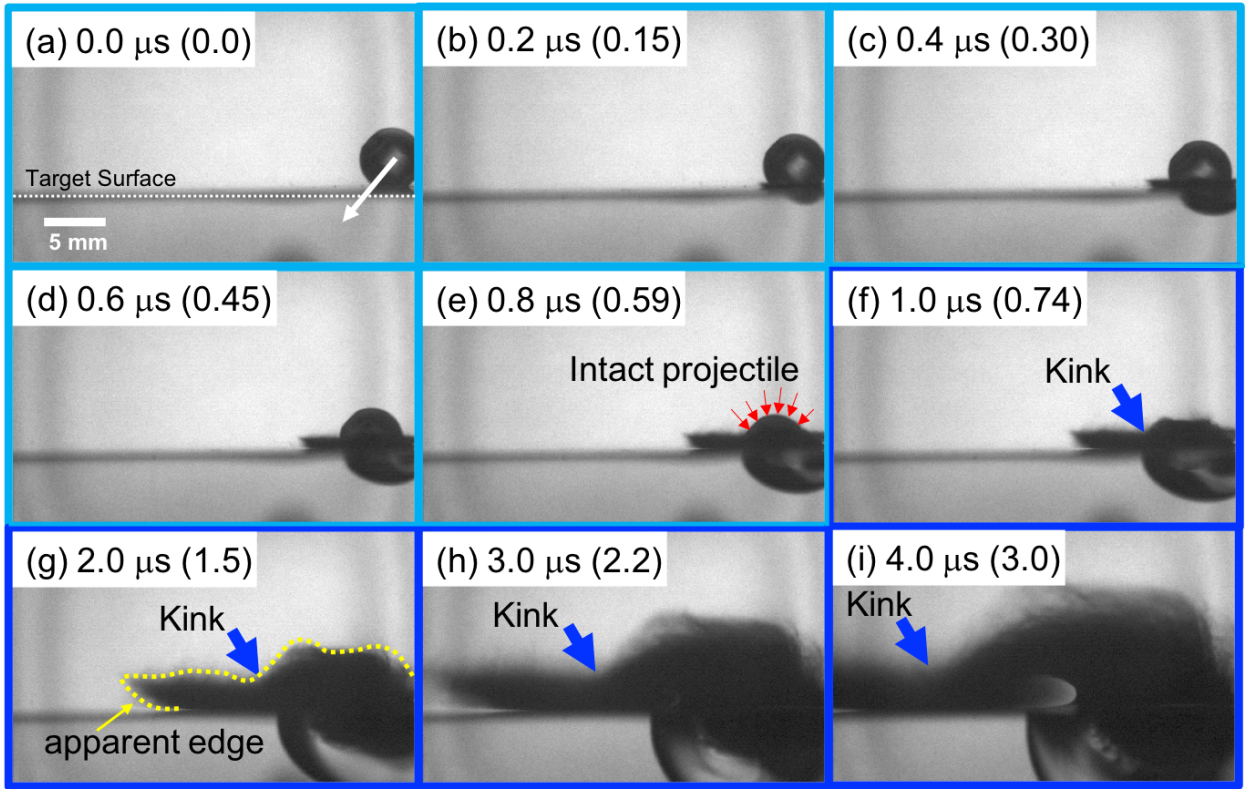

**Figure 4.** High-speed images for the oblique impact at an impact velocity of 3.56 km s$^{-1}$. The time following initial contact is indicated in each panel. Note that the time intervals between images in panels (a)–(f) and those in panels (f)–(i) are different. The numbers in parentheses indicate the scaled time, $t/t_s$, which is the ratio of the real time, $t$, to the characteristic time for projectile penetration, $t_s$. The kink structure is indicated by the blue arrows. The dotted yellow line in panel (g) shows an example of the apparent edge of the ejecta (see Section 3.2).

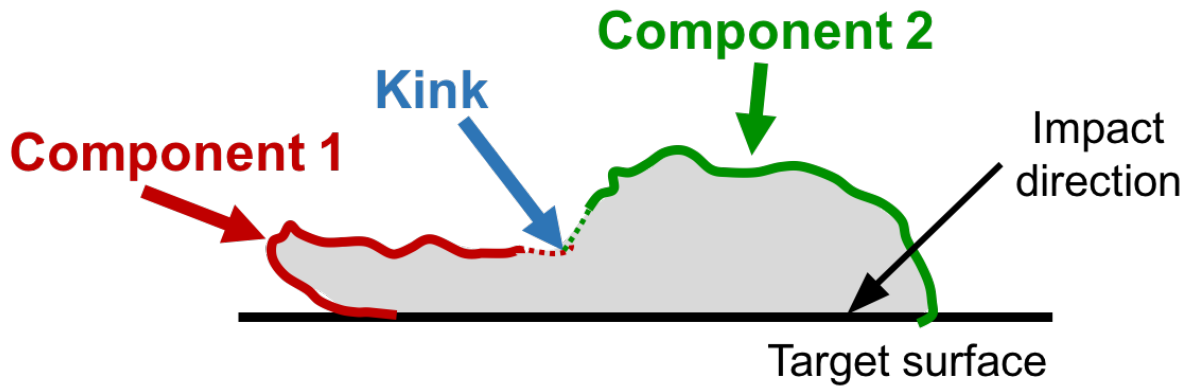

**Figure 5.** Schematic diagram of the ejecta. The ejecta consists of two components (see Section 3.1) and a kink structure is created between the two components.

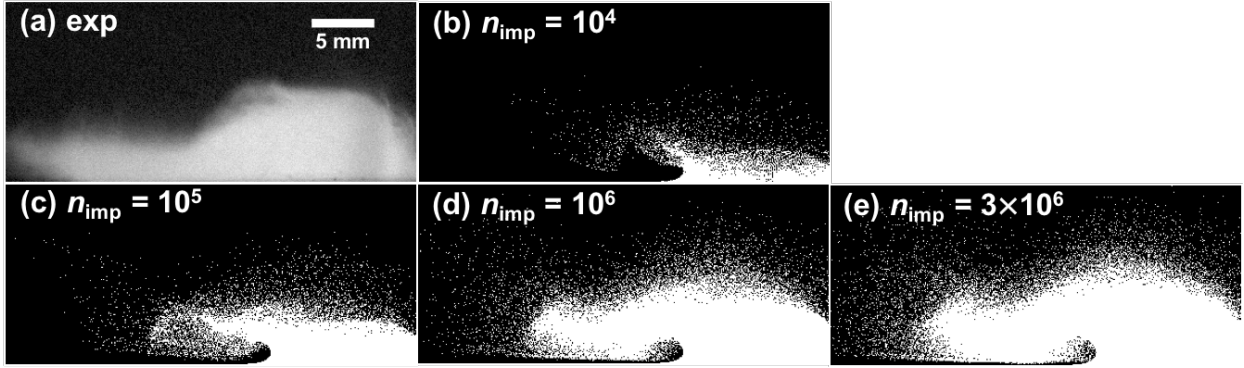

**Figure 6.** Processed images of the SPH simulations in panels (b)–(e) at an impact velocity of 3.56 km s$^{-1}$ for the oblique impact and for different $n_{\mathrm{imp}}$, as well as the subtracted image of the laboratory experiment, shown in panel (a). The bottom of each panel corresponds to the target surface. The time in all images is 3.0 μs (2.2 $t_s$) after the impact. The image of the laboratory experiment shown here was obtained by subtraction of the background from the raw image. The horizontal bar in panel (a) indicates the length scale.

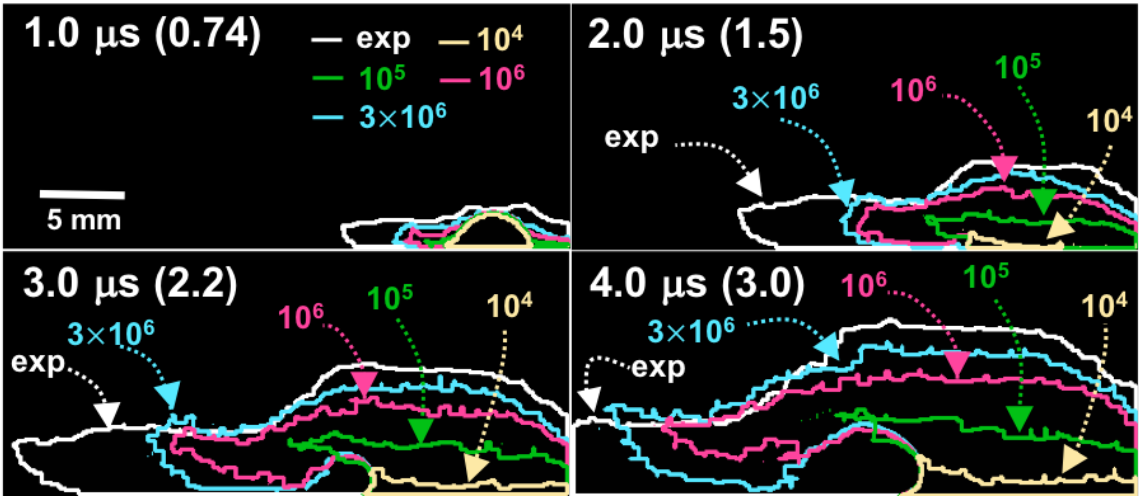

**Figure 7.** Apparent edge of the ejecta after image processing for the oblique impact at different times. The numbers in parentheses reflect the characteristic time for projectile penetration. Colors correspond to the color-coded result of the experiment and simulation for different $n_{\mathrm{imp}}$.

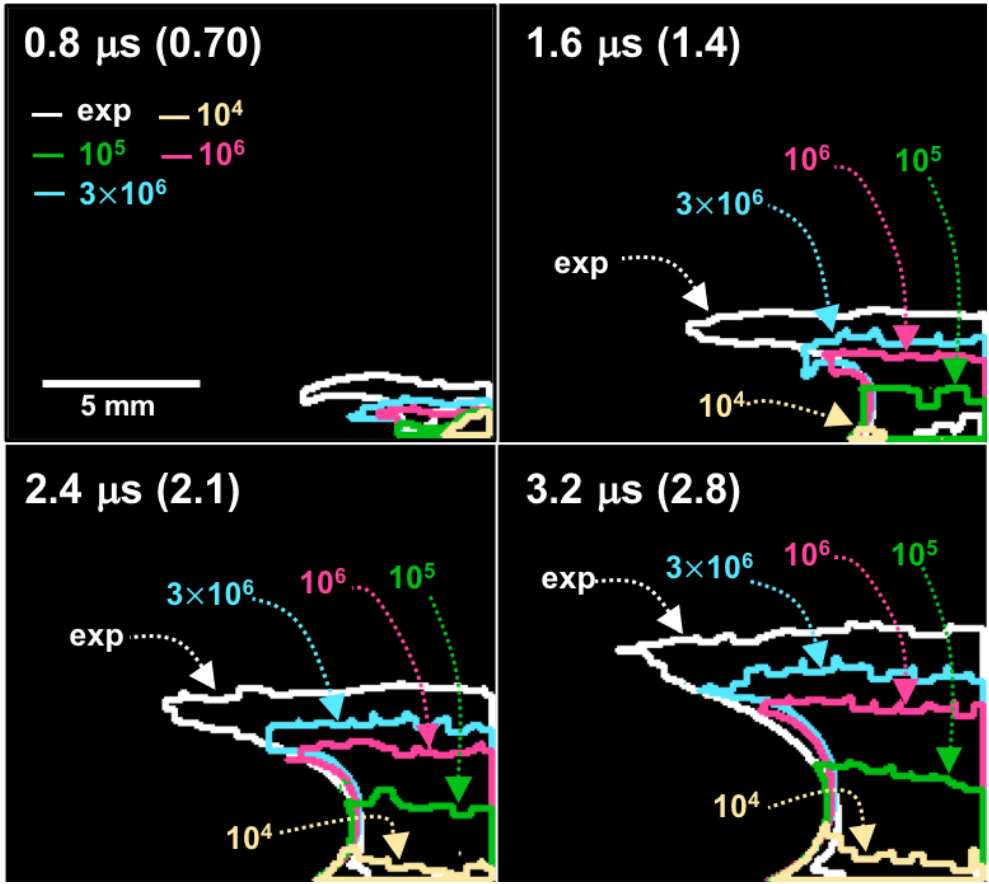

**Figure 8**. Apparent edge of the ejecta after image processing for the vertical impact at different times. Numbers in parentheses and colors correspond to the characteristic time and color-coded results for the apparent edge, respectively, as in Figure 7. Note that only half of the images of the ejecta edges are shown here because of the axial symmetry of the shape of the ejecta.

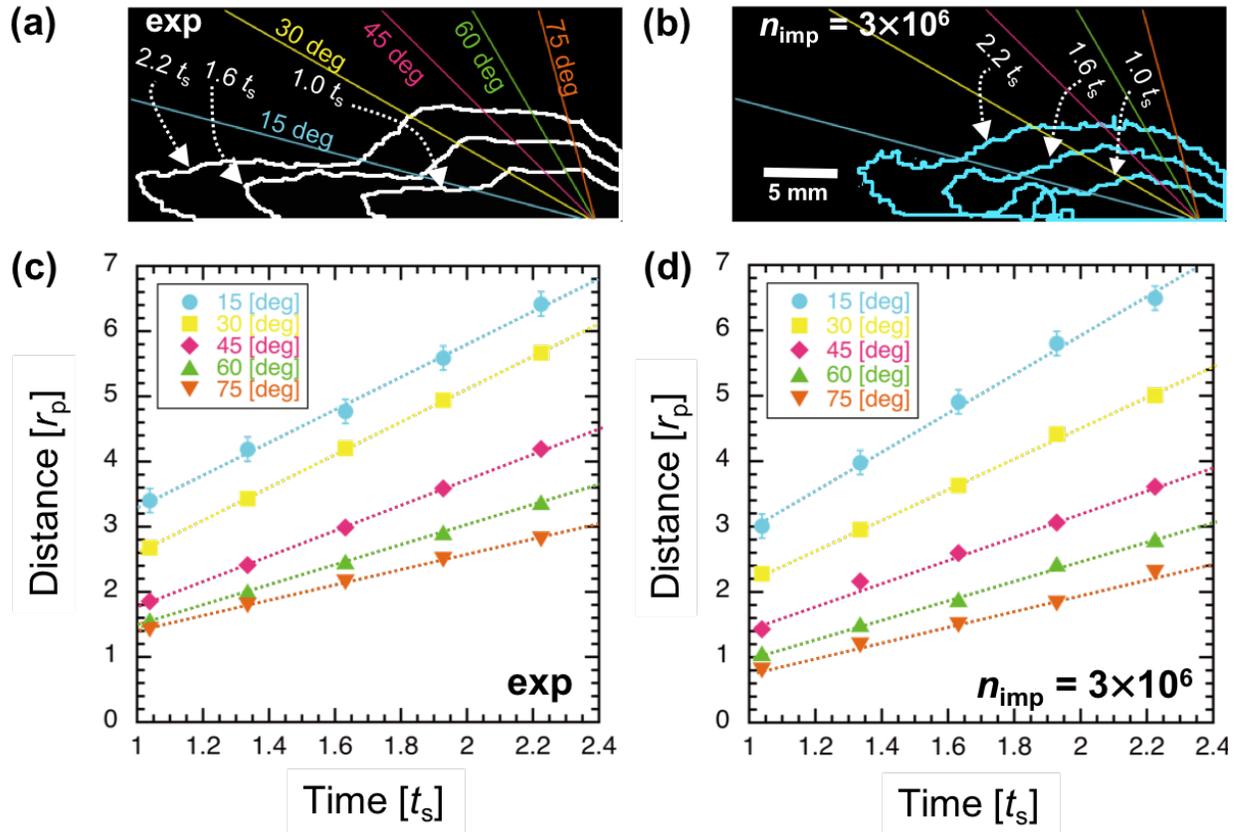

**Figure 9.** Distance between the impact point and apparent edges for different angles and as a function of time. The apparent edges for different times are represented in panels (a) and (b). The colored lines show the lines from the impact point for different angles from the target surface. Panels (a) and (c) show the results from the laboratory experiment, while panels (b) and (d) show the results from the simulation for $n_{imp} = 3\times10^6$. The dotted lines in panels (c) and (d) are the best-fitting linear functions to the five data points for each angle.

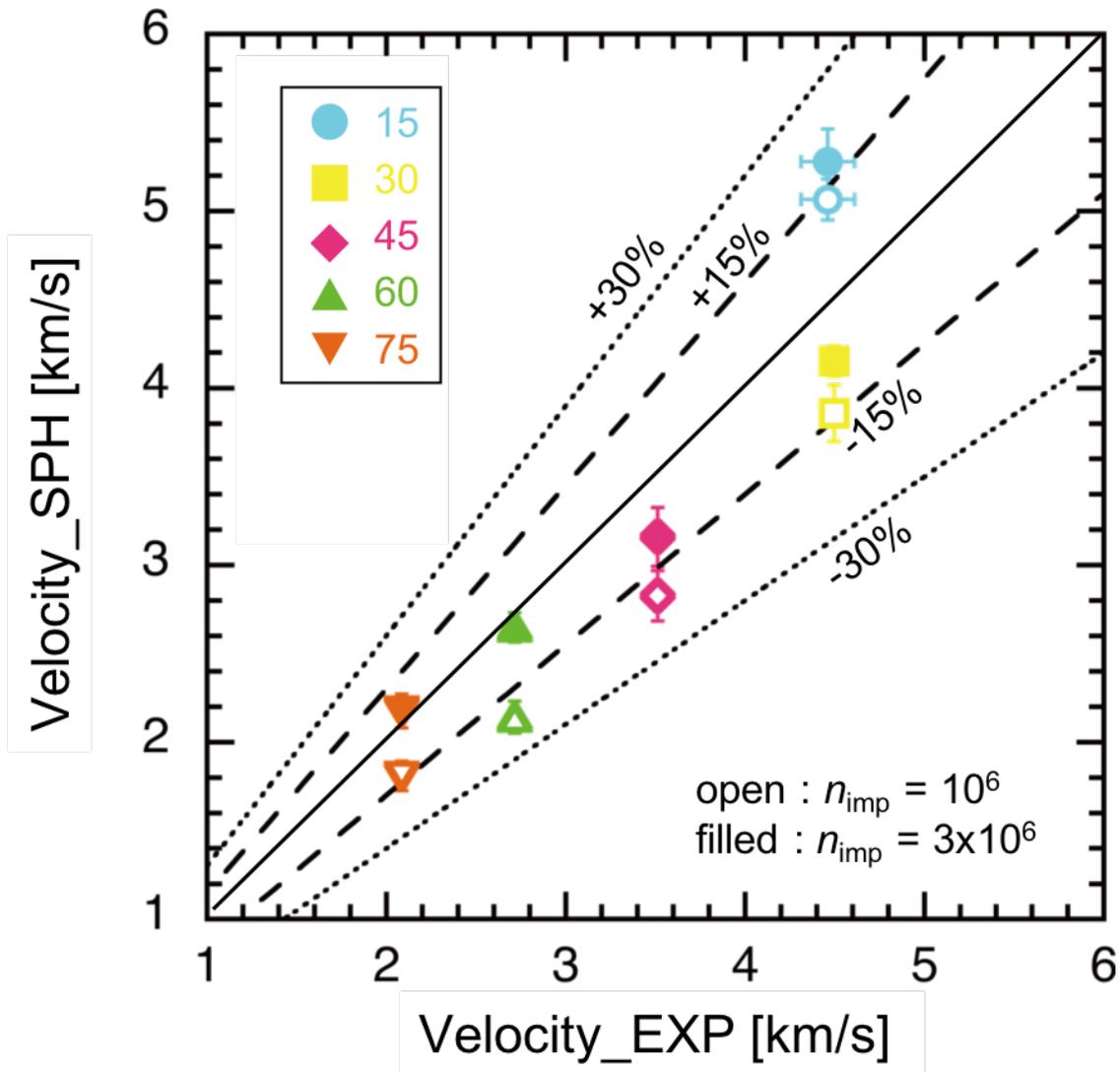

**Figure 10.** Comparison of the moving velocities of the apparent edge between the laboratory experiment and SPH simulation for different angles. The open and filled symbols show the results from the simulations for $n_{imp} = 10^6$ and $3 \times 10^6$, respectively. The numbers next to the symbols represent the angles from the target surface shown in Figure 9.

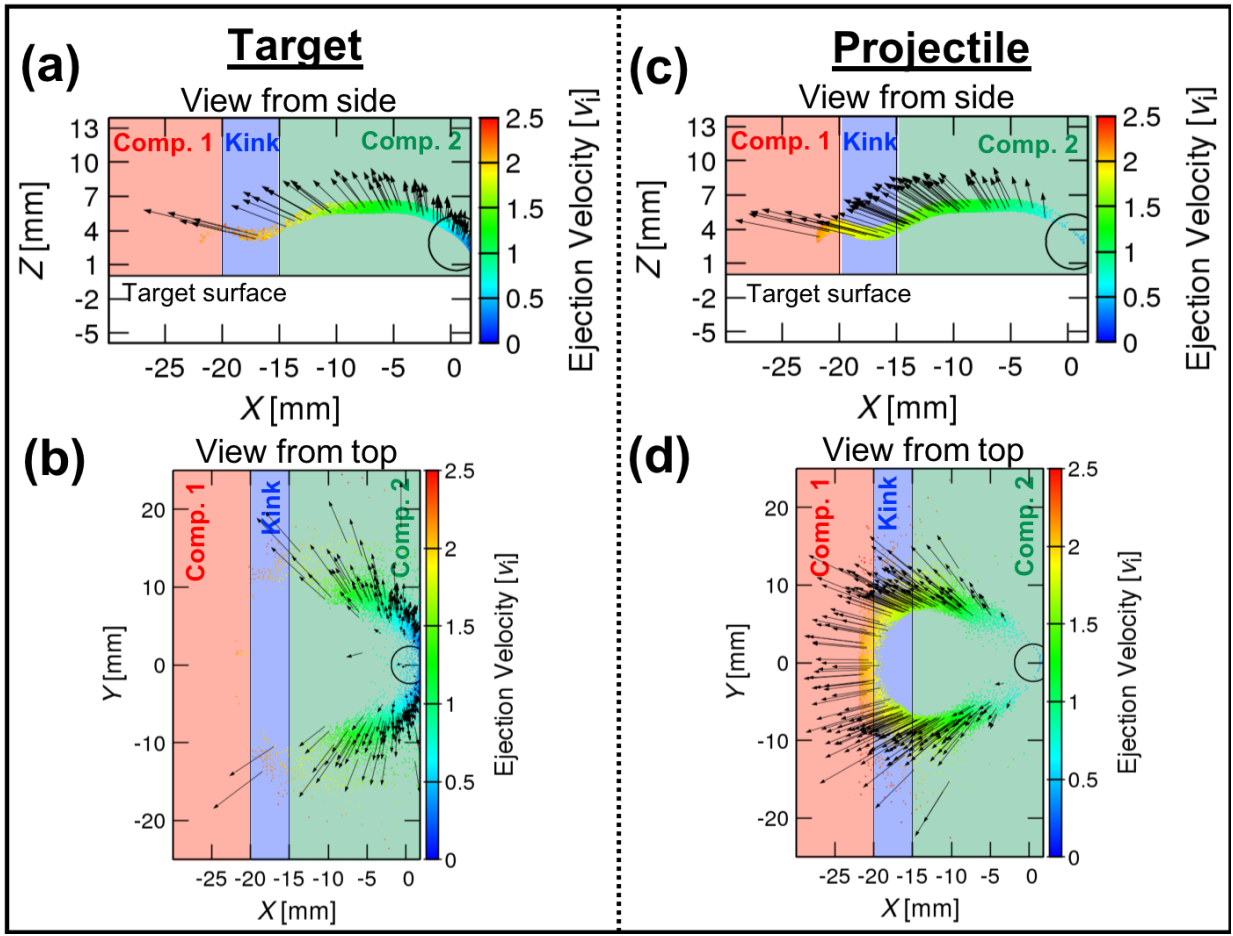

**Figure 11.** Spatial distributions of the extracted SPH particles at $t = 3.0$ μs ($t = 2.2\ t_s$). The SPH particles are color-coded depending on the ratios of the ejection velocities to the impact velocity. (a) Target materials in the $X$–$Z$ plane; (b) target materials in the $X$–$Y$ plane; (c) projectile materials in the $X$–$Z$ plane; (d) projectile materials in the $X$–$Y$ plane. The semi-transparent red, blue, and green areas correspond to the areas of component 1, the kink, and component 2, respectively. The arrows represent the velocity vectors of the particles' ejection velocities. The initial positions of the projectiles are indicated by circles.

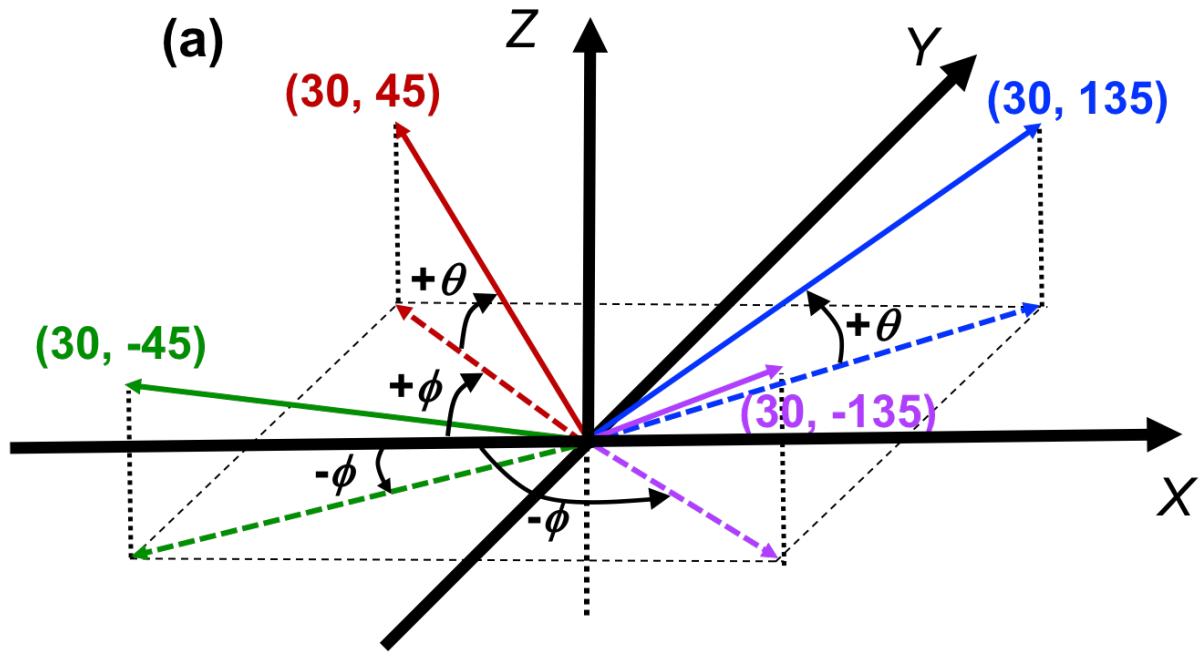
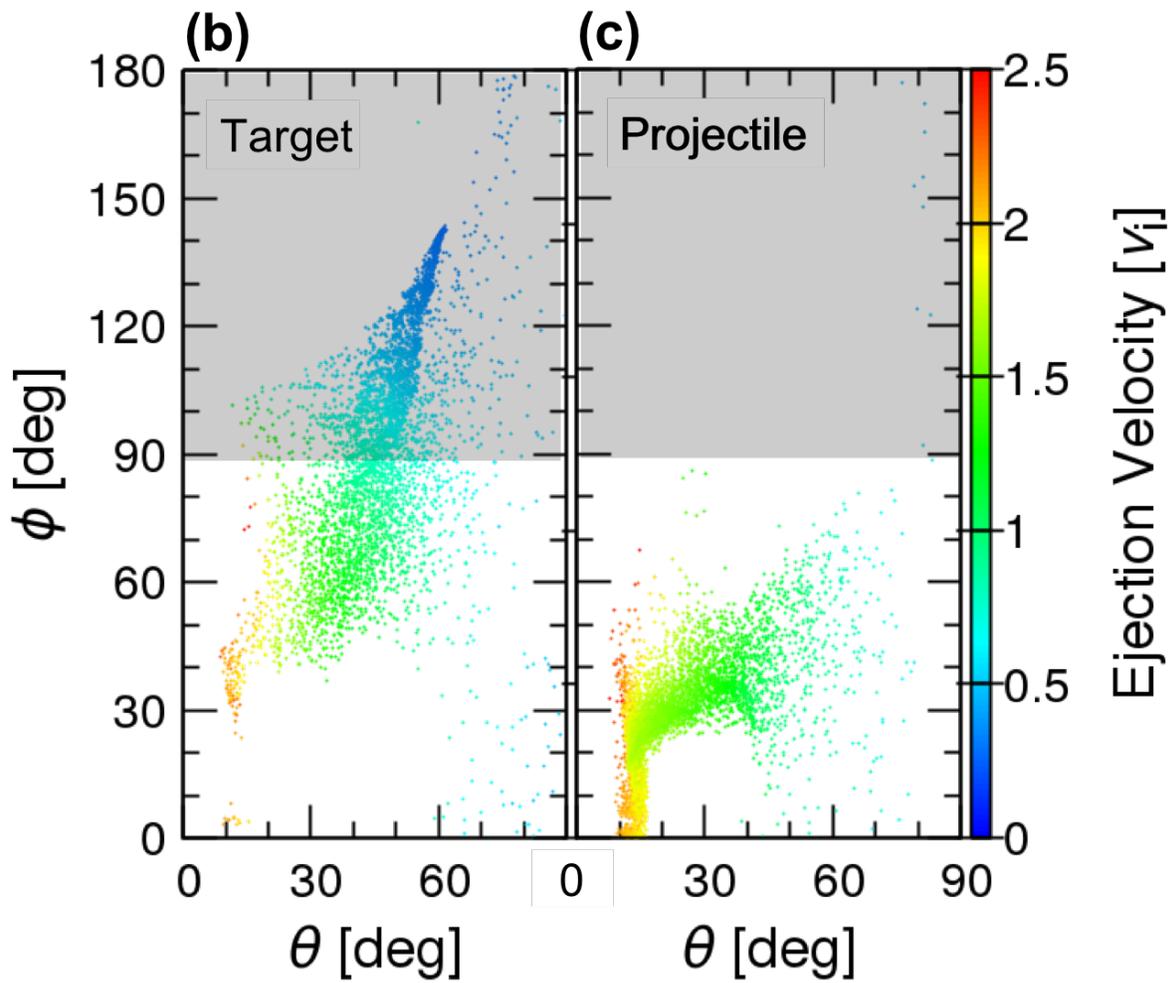

**Figure 12.** (a) Definition of the elevation angle, $\theta$, and azimuth angle, $\phi$. Examples for $(\theta, \phi) =$ (30, 45), (30, –45), (30, 135), and (30, –135) [in degrees] are shown. (b, c) Ejection velocity distributions of the extracted SPH particles at $t = 3.0$ μs ($t = 2.2\ t_s$) as a function of $\phi$ and $\theta$. The SPH particles from the target materials [panel (b)] and from the projectile [panel (c)] are color-coded by their ratios of the ejection velocities to the impact velocity, as in Figure 11. The ejection velocity to the upstream side of the impact is shown in the gray area.

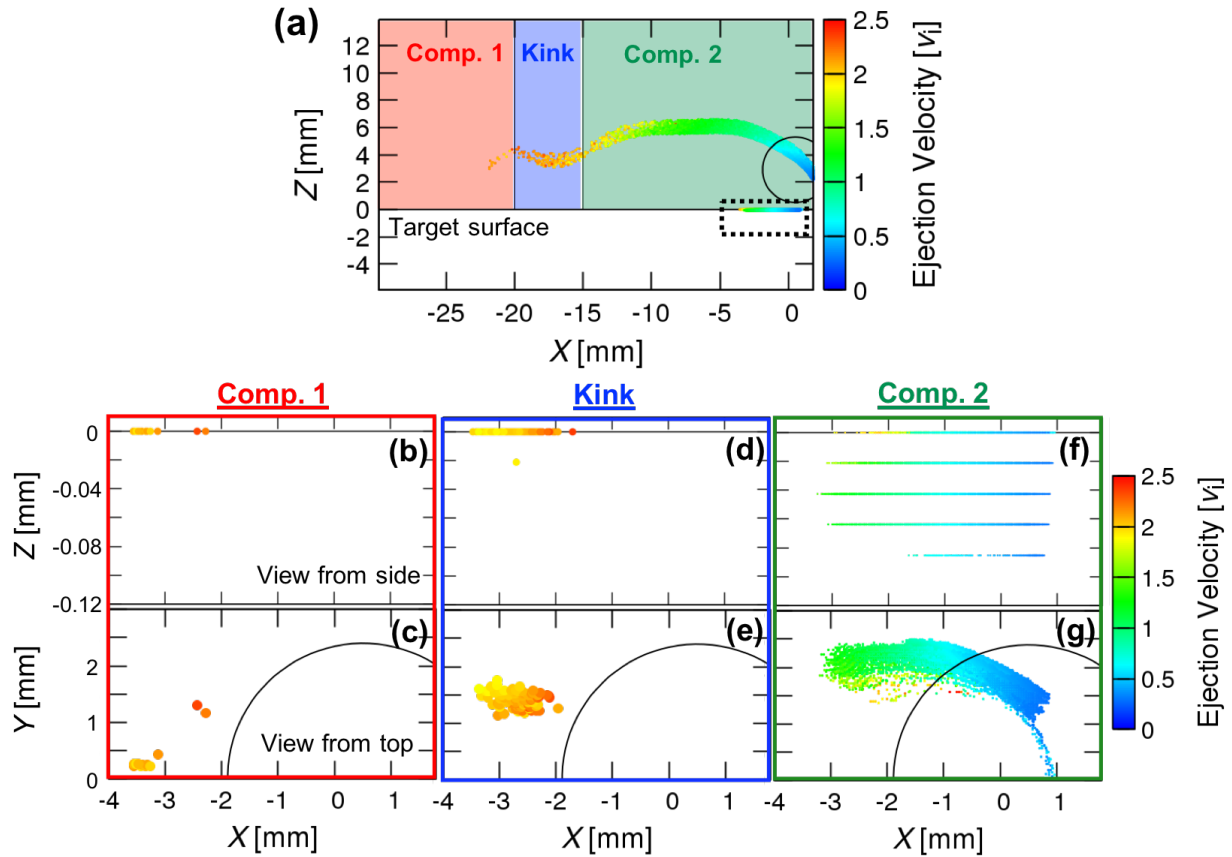

**Figure 13.** Initial positions of the extracted SPH particles in the target materials shown in Figure 11. The SPH particles are color-coded by their ratios of the ejection velocities to the impact velocity. The initial positions of the projectiles are shown as circles. (a) Initial positions of the extracted SPH particles in the *X–Z* plane with the positions of the particles shown at $t = 3.0$ μs ($t = 2.2\ t_s$). Enlarged views of the particles in the dotted rectangle are shown in panels (b)–(g). Panels (b) and (c) show the initial positions of the extracted SPH particles from component 1 in the *X–Z* and *X–Y* planes, respectively. Panels (d) and (e) show the initial positions of the extracted SPH particles in the kink in the *X–Z* and *X–Y* planes, respectively. Panels (f) and (g) show the initial positions of the extracted SPH particles from component 2 in the *X–Z* and *X–Y*

planes, respectively. The particles in panels (b)–(e) are shown using larger symbols for display reasons.

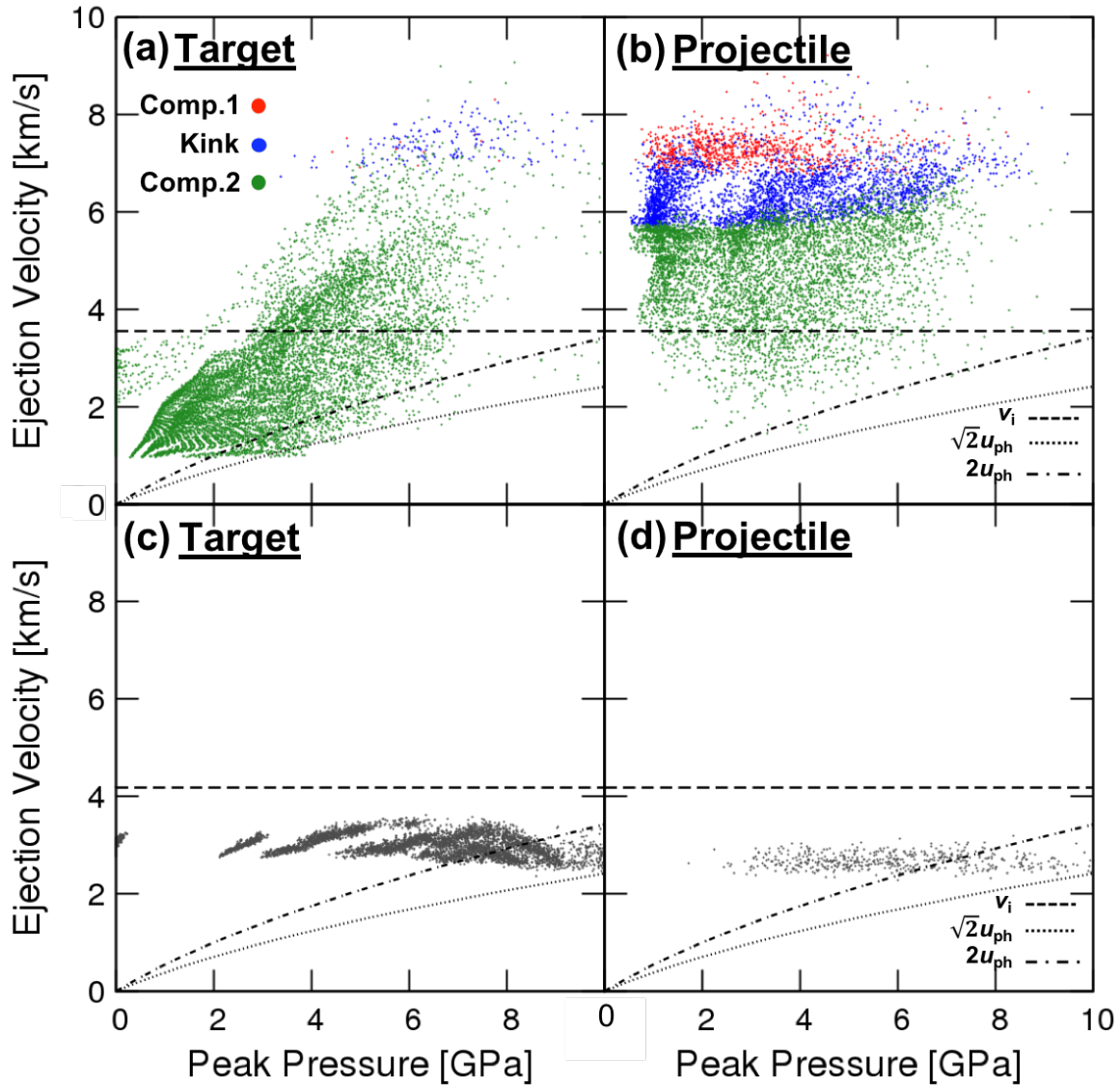

**Figure 14.** Ejection velocities as a function of the peak pressure of the extracted particles. The spatial distributions of the particles are shown in Figures 11 and C5. Panels (a) and (b) show the results of an oblique impact at 3.56 km s$^{-1}$ and 45 degrees. Panels (c) and (d) show the results of a vertical impact at 4.18 km s$^{-1}$. We divided the ejecta particles from the target and projectile into different panels labeled "Target" and "Projectile". The red, blue, and green symbols correspond to particles associated with component 1, the kink, and component 2 (see Section 4.1), respectively. The horizontal dashed line shows the impact velocity. The dotted and dash–

dotted lines show the lines where the ejection velocity is equal to $\sqrt{2}u_{\mathrm{ph}}$ and $2u_{\mathrm{ph}}$, respectively, calculated using the 1D impedance match solution (e.g., Melosh, 1989). The shock Hugoniot parameters used in this calculation were taken from Marsh et al. (1980).

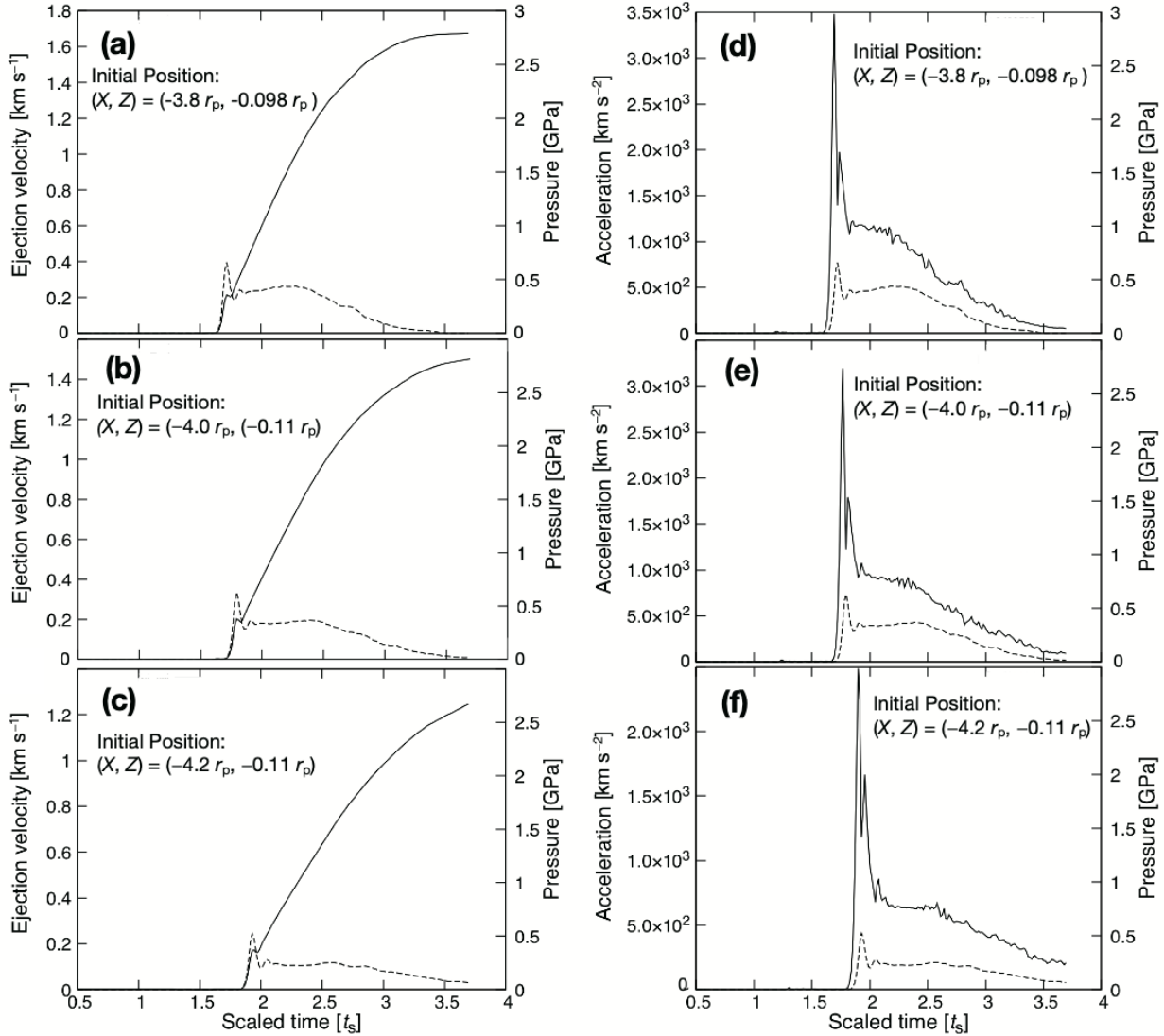

**Figure 15.** (a–c) Temporal variations of the particle velocity (solid line; left $Y$ axis) and pressure (dashed line; right $Y$ axis) for selected SPH particles experiencing post-shock acceleration. (d–f) Temporal variations of the acceleration (solid line; left $Y$ axis) and pressure (dashed line; right $Y$ axis) for the same SPH particles. The values in parentheses in each panel reflect the initial ($X$, $Z$) positions of the particles in units of the projectile radius, $r_{\mathrm{p}}$.

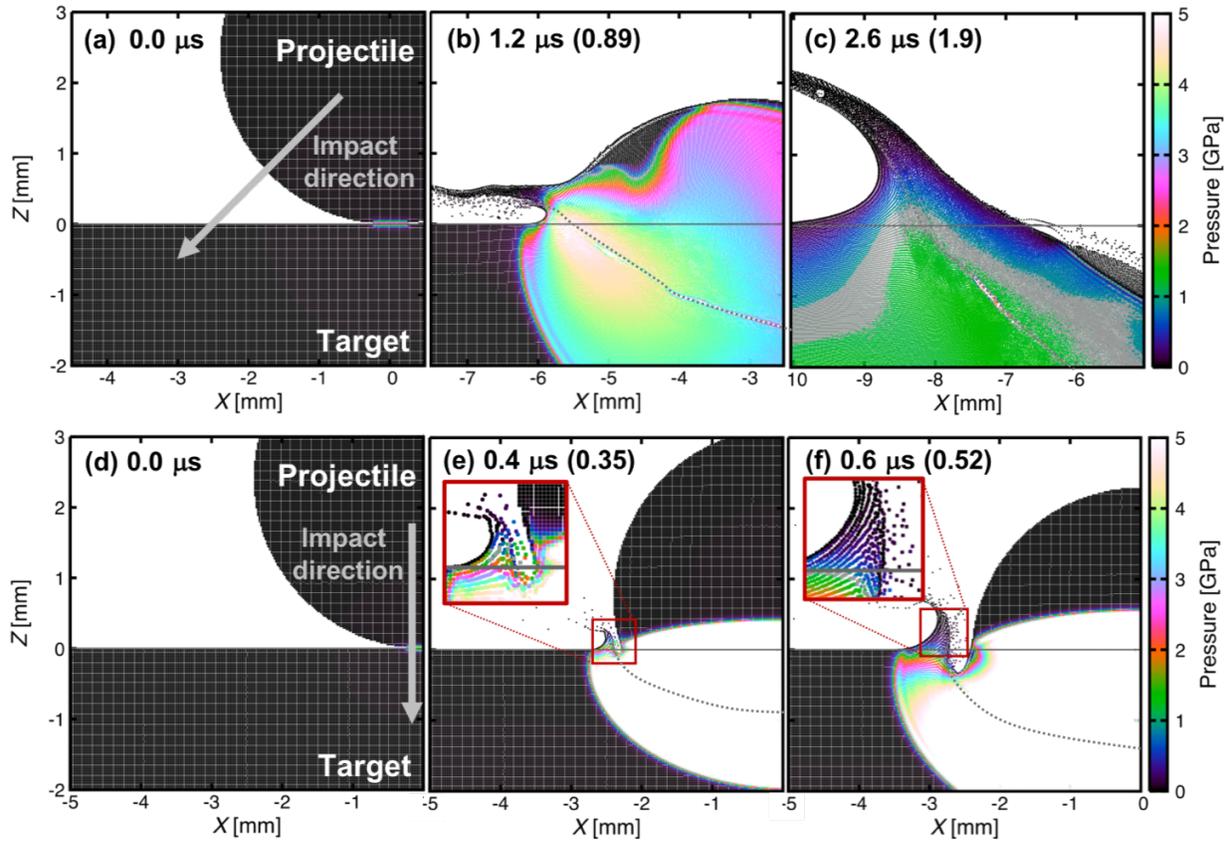

**Figure 16.** Snapshots of close-up views in the vicinity of the impact point. SPH particles are color-coded depending on their temporal pressures (i.e., not their peak pressure). Particles with pressures around 1 GPa are colored gray. The characteristic time is shown in parentheses. The boundaries between the projectiles and targets are shown in gray dotted lines. (a–c) Snapshots for an impact velocity of 3.56 km s$^{-1}$ and an impact angle of 45 degrees. (d–f) Snapshots for an impact velocity of 4.18 km s$^{-1}$ and an impact angle of 90 degrees. The insets in panels (e) and (f) are enlarged images of the root of the ejecta. The specific values used in Eqs (2) and (3) were measured from panels (c) and (e), respectively.

Supporting Information for

Impact Ejecta near the Impact Point Observed using Ultra-high-speed Imaging and SPH Simulations, and a Comparison of the Two Methods

Takaya Okamoto[1†], Kosuke Kurosawa[2], Hidenori Genda[3], and Takafumi Matsui[2]

[1] Institute of Space and Astronautical Science, Japan Aerospace Exploration Agency, Japan

[2] Planetary Exploration Research Center, Chiba Institute of Technology, Japan

[3] Earth–Life Science Institute, Tokyo Institute of Technology, Japan

**Contents of this file**

 Text S1 to S6
 Figures S1 to S16

**Introduction**

 This document includes pre-processed images (Text S1; Figure S1), descriptions of the detail image processing (Text S2; Figures S2–S5), the moving velocities of the apparent edge of the oblique impact for an impact velocity of 5.04 km s$^{-1}$ (Text S3; Figures S6–S7), the results and the velocity distributions of the vertical impact for an impact velocity of 4.18 km s$^{-1}$ (Text S4.1–S4.2; Figures S8–S14), the mass distribution of the high-speed ejecta at the oblique impact (Text 5; Figure S15), and the accuracy of the EOS model in the simulations (Text 6; Figure S16).

**Text S1. Snapshots in the SPH simulation before Data Analysis and Image Processing**

 We show a large number of post-processed images above the target surface in the main text in order to quantitatively compare the results of the high-speed ejecta in the numerical simulations with those of the laboratory experiments. Here, in Text S1, we show the pre-processed images from the SPH simulations for $n_{imp} = 3 \times 10^6$, before the data analysis described in Section 2.3 and image processing described in Text S2. Figure S1(a)–(c) shows images of a cross-section in the projectile's trajectory plane. Only particles on the cross-section in the *X–Z* plane for $Y < \pm 0.022$ mm, where the two layers are closest to the $Y = 0$ plane, are shown. A growing crater can be observed during projectile penetration. Figure S1(d)–(f) shows images of ejecta above the surface projected onto the *X–Y* plane. Only particles above the target surface ($Z > 0.0$ mm) and with their positive particle velocity component in the *Z* direction were displayed in order to observe the ejecta (i.e., particles of the projectile with a negative velocity component in the *Z* direction during penetration were excluded). Particles moving at high ejection velocities of twice the impact velocity are distributed on the leading side, whereas many particles with low ejection velocities and some particles with velocities comparable to the impact velocity immediately after the impact are ejected from the upstream side of the impact.

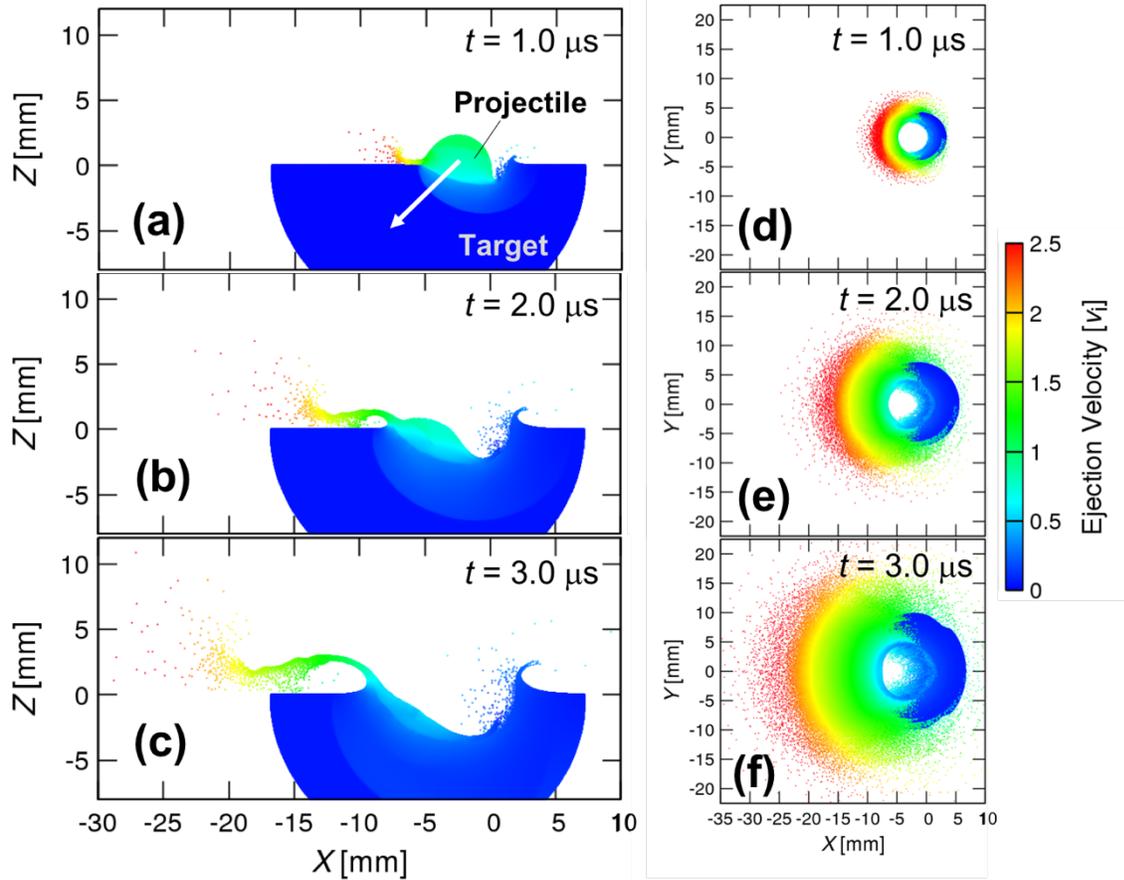

**Figure S1.** Pre-processed images of the SPH simulations at an impact velocity of 3.56 km s$^{-1}$ for the oblique impact for $n_{\text{imp}}= 3\times10^6$. The SPH particles are color-coded depending on the ratios of the ejection velocities to the impact velocity. Panels (a)–(c), and (d)–(f) show the cross-section images in the projectile trajectory plane, and the images of ejecta projected to the $X$-$Y$ plane at time every 1μs, respectively.

**Text S2. Image Processing to Extract the Apparent Edge of the Ejecta**

      The details of the image processing used to produce Figures 7 and 8 in the main text are described in this Text S2. The image processing pertaining to the results of the laboratory experiment and SPH simulations for an impact velocity of 3.56 km s$^{-1}$ is mainly presented here. The same procedure was also applied to the results pertaining to the oblique impact for an impact velocity of 5.04 km s$^{-1}$ as well as for the vertical impact. The processing was conducted using OpenCV (e.g., Bradski & Kaehler, 2008). To determine the location of the apparent edge of the ejecta, we applied a binarization technique to the images and then extracted the outlines of the curtain. Although the apparent edge determined by the binarization technique may not be the exact edge of the ejecta, the edge must be determined by a certain threshold value in the counts on the image sensor, because the location of the edge in the images captured

by the camera is determined by the balance between the optical depth of the ejected particles along the line of sight and the illumination from the light source.

We analyzed the high-speed images from the laboratory experiment shown in Figure 4 in the main text and the processed images of the SPH simulations shown in Figure 6(b)–(e) in the main text using the same method. However, the observed count on the detector of the camera was used to process the high-speed images, while the numbers of SPH particles in each cell were treated as pixel values while processing the images of the SPH simulations. Note that background images were subtracted from the raw high-speed images. The background images were taken before the relevant shot under the same filming conditions. Regions above the target surface in the images were cut out in advance. In addition, for the vertical impact, regions at $X > 0$ in the images were cut out because the morphology of the ejecta was symmetrical about the $Z$ axis. To obtain the inside region of the edge of the ejecta, a binarization technique was used (i.e., pixel values after binarization, $B(i,j)$, were assigned):

$$B(i,j) = \begin{cases} 255, & I(i,j) > \Gamma \\ 0, & I(i,j) \leq \Gamma \end{cases}, \qquad (S1)$$

where $i, j, I(i,j)$, and $\Gamma$, are column and row numbers, pixel values or number of SPH particles, and a threshold of the pixel value or number of SPH particles applied for binarization, respectively. Pixel values range from 0 (black) to 255 (white) for an 8-bit image. We refer to the images after the binarization procedure as "EXP" images for the laboratory experiments and "SPH" images for the SPH simulations. Figures S2 and S3 show the EXP and SPH images for $n_{imp} = 3 \times 10^6$, respectively. A number of isolated white pixels for small values of $\Gamma$ are notable in Figures S2(b) and S3(a). The number of isolated white pixels and areas of the ejecta, which correspond to the largest continuous white pixel group (hereafter, referred to as "block"), decrease with increasing $\Gamma$. The morphologies of the ejecta could not easily be extracted in Figures S2(f) and S3(e–f) for high values of $\Gamma$, compared with the morphology in the subtracted image of the experiment shown in Figure S2(a). Since the size of each block depends on $\Gamma$, we have to choose an appropriate value of $\Gamma$ to extract the edge. We determined a set of $\Gamma$ for both the EXP and SPH images at a given time through a trial-and-error procedure based on two criteria: (1) the isolated white pixels are mostly removed and (2) both images exhibit a good correlation. A normalized cross-correlation (NCC) coefficient, $R_{ncc}$, between the EXP and SPH images was used to determine the correlation, defined as (e.g., Bradski & Kaehler, 2008)

$$R_{ncc} = \frac{\sum_{i,j}[B_{exp}(i,j) \cdot B_{sph}(i,j)]}{\sqrt{\sum_{i,j}[B_{exp}(i,j)]^2 \cdot \sum_{i,j}[B_{sph}(i,j)]^2}}, \qquad (S2)$$

where $B_{exp}$ and $B_{sph}$ are pixel values in the EXP and SPH images, respectively. Finally, we chose the $\Gamma$ =19 set for the EXP images and $\Gamma$=3 for the SPH images. This combination yields high value of $R_{ncc}$ = 0.80. The $\Gamma$=32 set for the EXP images and $\Gamma$=3 for the SPH images, as well as the $\Gamma$=59 set for the EXP images and $\Gamma$=3 for the SPH images, were chosen for the oblique impact for an impact velocity of 5.04 km s$^{-1}$ and for the vertical impact, respectively. The corresponding values of $R_{ncc}$ are 0.79 and 0.85, respectively. Note that the locations of the edges in the low-resolution SPH simulations appear similar to those in the high-resolution simulations for large binarization thresholds (see Figures 6 and 7 in the main text and Figure S3). This apparent resemblance indicates that our SPH code scales well with the number of SPH particles (i.e., the hydrodynamic motion of one SPH particle for $n_{imp} = 10^5$ is well characterized by the mean value of 10 SPH particles for $n_{imp} = 10^6$). This result supports the validity of our numerical code.

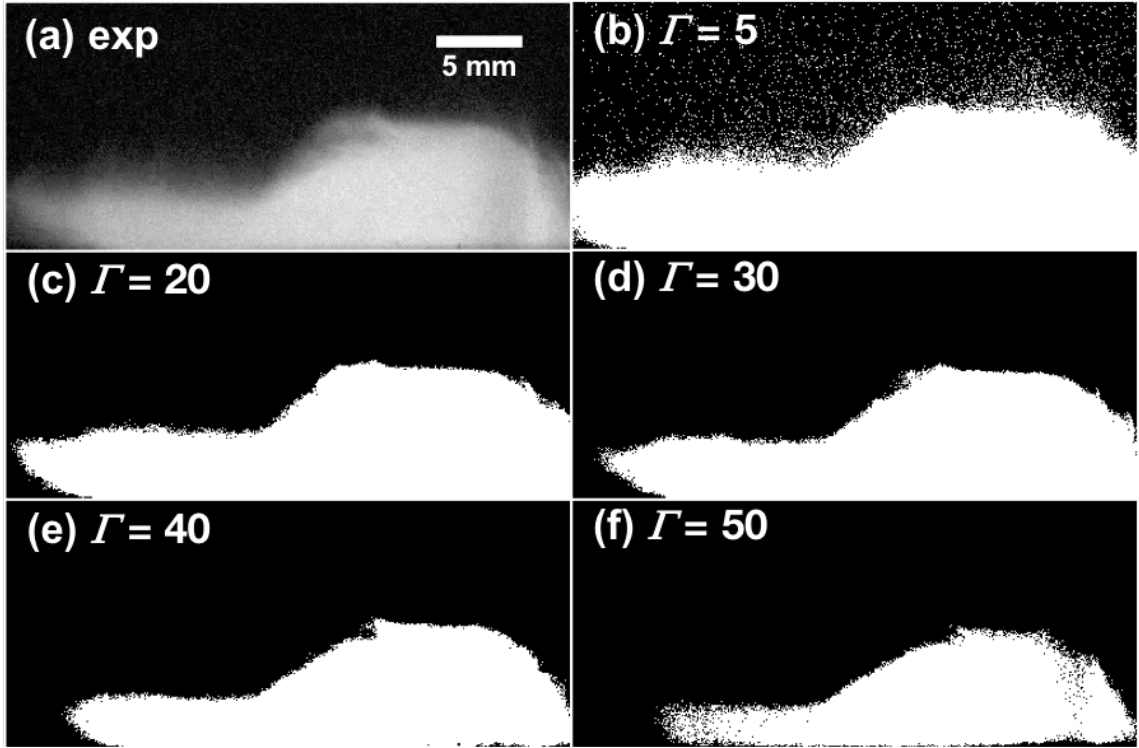

**Figure S2.** Examples of images from the experiment (EXP) at $t = 3.0$ μs for a given binarization threshold, $\Gamma$—panels (b)–(f)—as well as the subtracted image of the impact experiment, shown in panel (a).

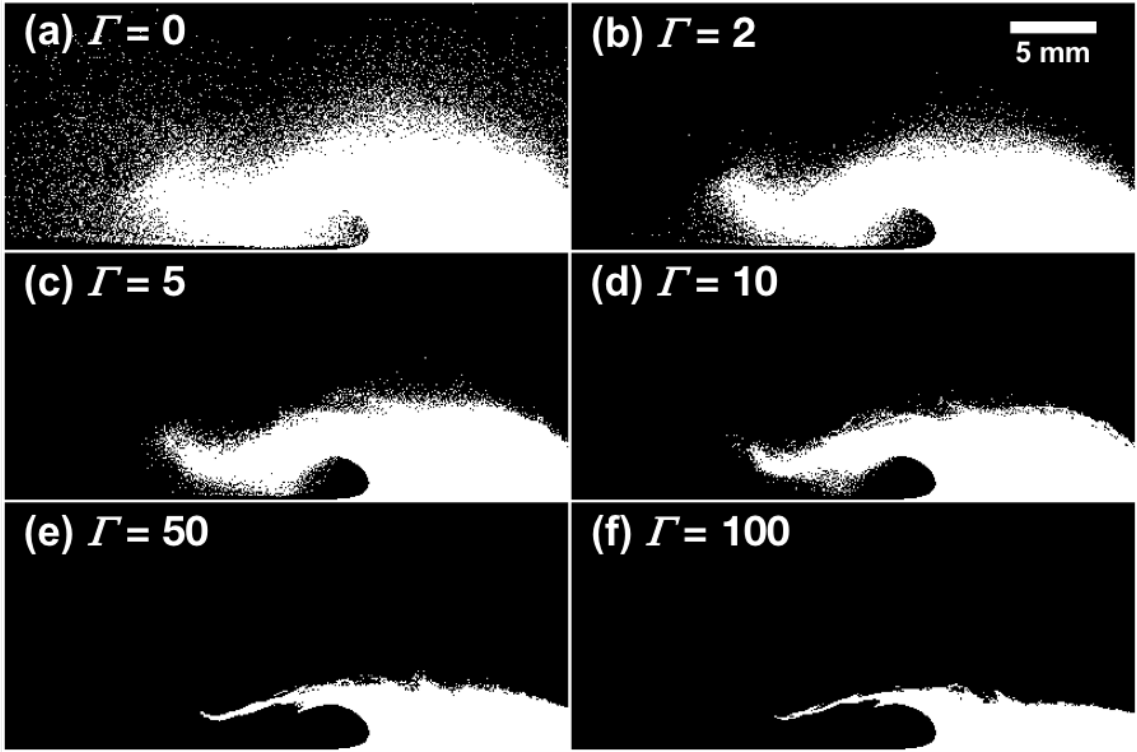

**Figure S3.** Examples of images from the simulations (SPH) at $t = 3.0$ μs for a given binarization threshold, $\Gamma$.

A small number of isolated white pixels still remain around the apparent edge of the ejecta in the binary images, as shown in Figure S4(a). If we extracted the outline of the block using an OpenCV contour-finding function in the binary images, the extracted edge exhibits a zig-zag shape and shows many spikes, which may be artifacts. Thus, morphological image processing operations, such as opening and closing operations, were conducted in preparation. Opening and closing operations manipulate the erosion and dilation processes to improve images (i.e., they are used to remove noise, isolate individual elements, and join disparate elements in an image). A closing operation based on a 3×3 kernel followed by an opening operation using the same kernel were applied to the binary images for the given thresholds. These operations enabled us to remove the artificial spikes and extract the smoother apparent edge of the ejecta. Figure S4 shows example images before and after the operations. Finally, we extracted the edges of the ejecta from the images after the operations, as shown in Figures 7 and 8 in the main text. The shape of the extracted edge after the operations was much smoother than before the operations, as shown in Figures S4(c) and S4(e).

Note that the moving velocity is not significantly affected by changing the threshold value for image binarization, although the location of the extracted edge depends on this value. Figure S5 shows the moving velocity of the apparent edge in the laboratory experiment as a function of the binarization threshold for the experiment, $\Gamma$. The moving velocities in the direction 45° from the target surface were measured as a representative direction, as well as for Figure 9 in the main text, showing that the moving velocities of the apparent edges do not depend significantly on the threshold value and that their values are within ~4% difference (between $\Gamma$ = 10 and 40), which is smaller than the 15% difference between the velocities in the experiments and numerical results.

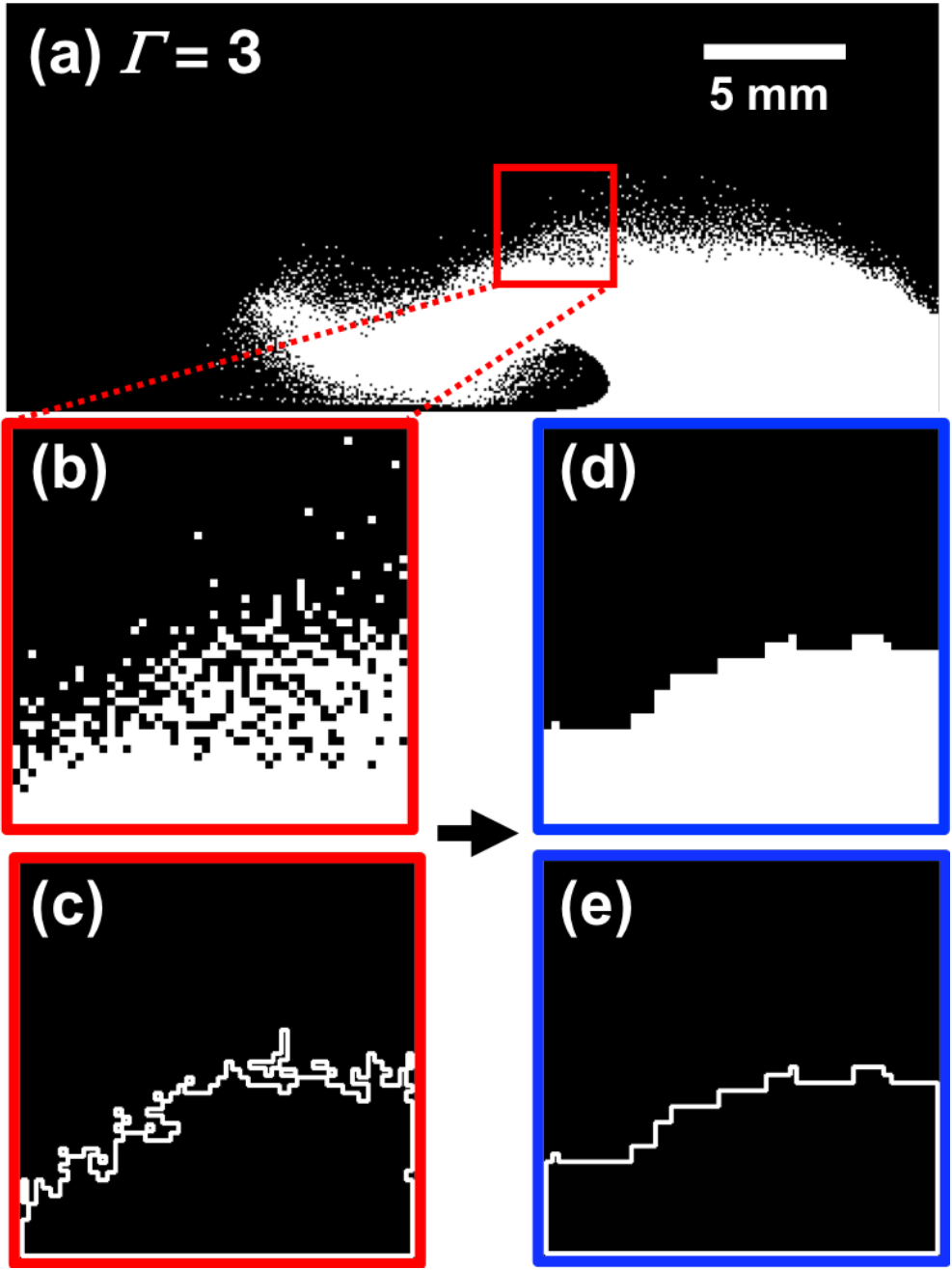

**Figure S4.** Examples of images before and after the closing and opening operations. (a) SPH images for $\Gamma$ = 3 before the operations. (b) Enlarged image of the red area in panel (a). A number of isolated white and black pixels remained around the edge of the ejecta. (c) Contour line of the largest block in panel (b). The shape of the extracted edge was rather artificial. (d) Image after application of the closing and opening operations in the same area as shown in panel (b). Isolated pixels around the edge were removed. (e) Contour line of the block in panel (d).

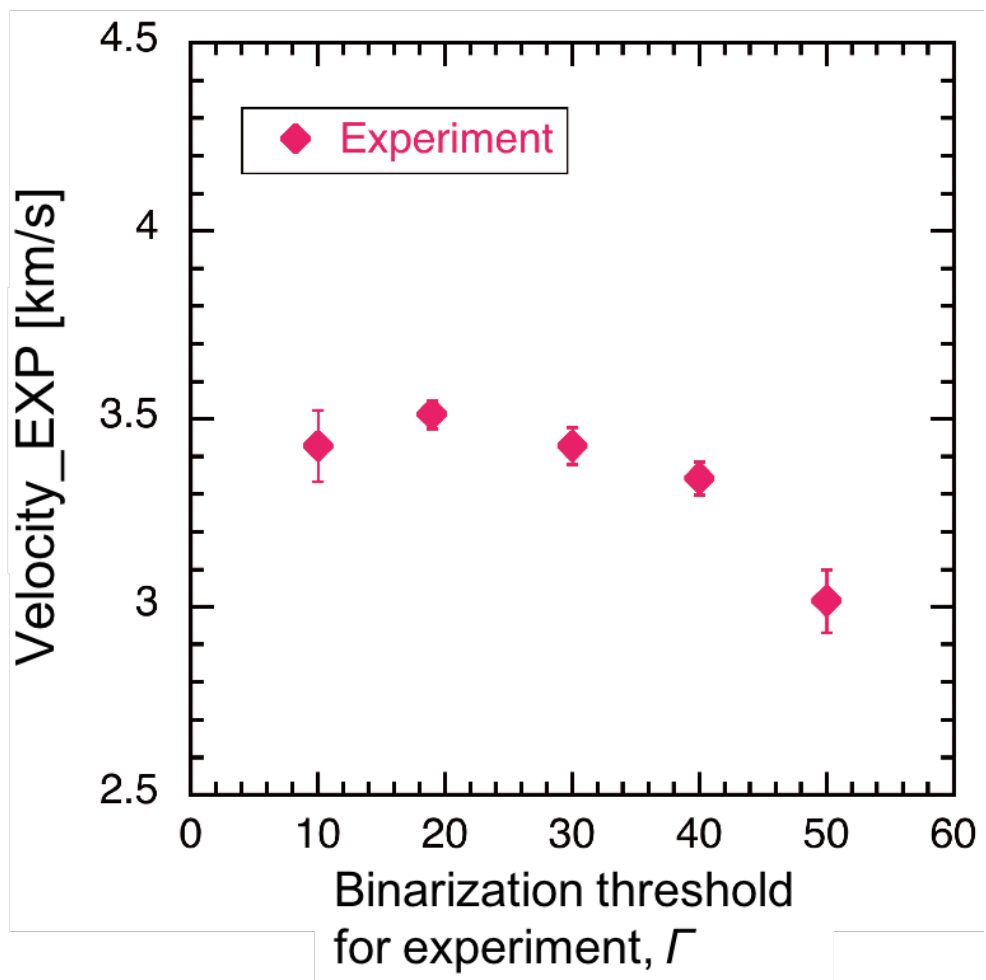

**Figure S5.** The moving velocity of the apparent edge in the laboratory experiment toward the direction of 45° from the impact point with respect to the target surface as a function of the binarization threshold for the experiment $\Gamma$.

**Text S3. Apparent Edge of the Oblique Impact for an Impact Velocity of 5.04 km s$^{-1}$**

We extracted the apparent edge of the ejecta using the same procedures as in Text 2 from a shot for an impact velocity of 5.04 km s$^{-1}$ and an angle of 45 degrees. Figures S6(a) and S6(b) show examples of the edges at $t$ = 1.1, 1.9, and 2.7 $t_s$ ($t$ = 1.0, 1.8, and 2.6 μs, respectively). The two components (components 1 and 2), which were observed for an impact velocity of 3.56 km s$^{-1}$, also appeared for the higher impact velocity of 5.04 km s$^{-1}$. Figures S6(c) and S6(d) show the distance between the impact point and apparent edges for different angles as a function of time. The edges' moving velocities for different angles, obtained from the slopes in Figures S6(c) and S6(d), are shown in Figure S7. The results show that the velocities in the simulations for $n_{imp} \geq 10^6$ are in agreement (i.e., ±30%) with the velocities in the laboratory experiment. Especially for $n_{imp} = 3 \times 10^6$, most velocities are in good agreement, to within ±15% (see the detailed description of the results in Section 3.2 in the main text).

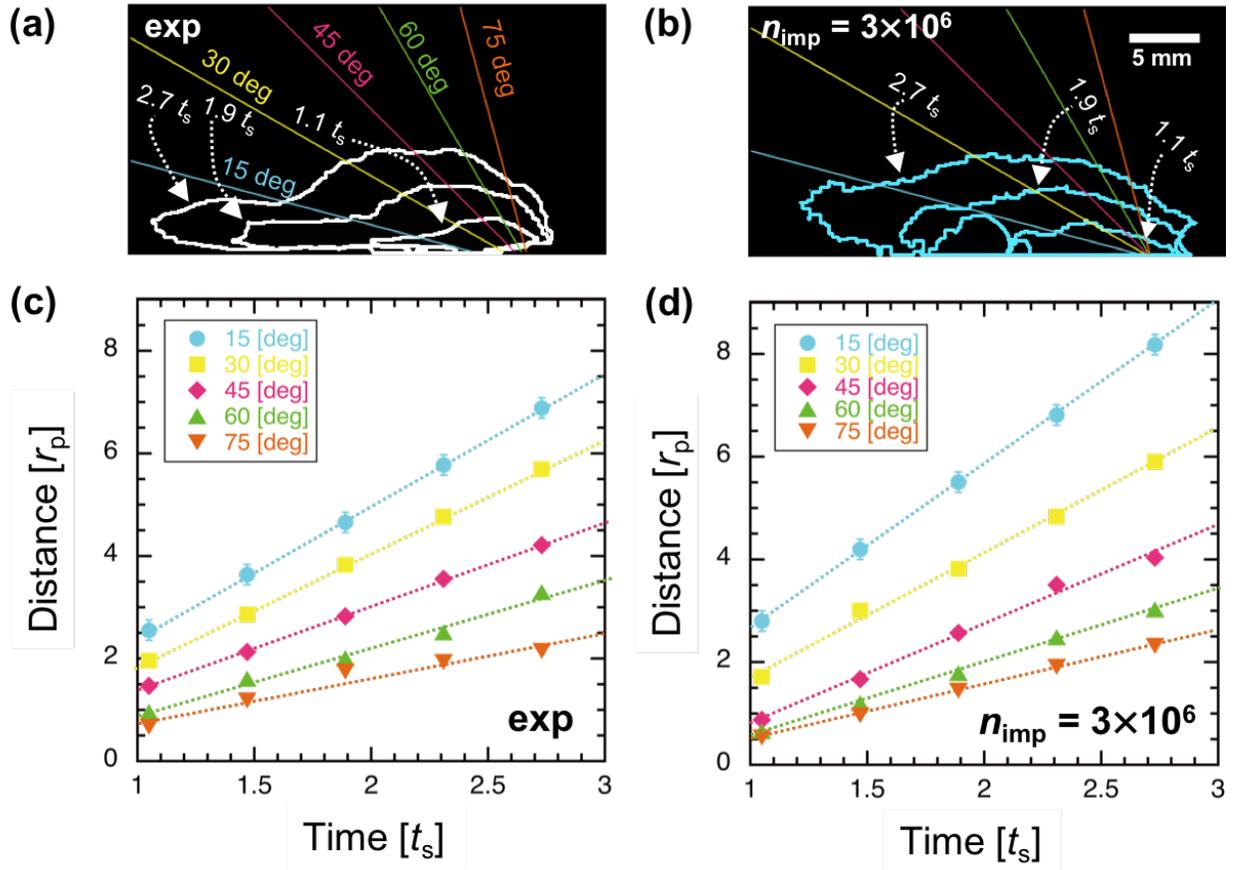

**Figure S6.** Distance between the impact point and apparent edges for different angles as a function of time. This figure shows the results of the shot for an impact velocity of 5.04 km s$^{-1}$ and an impact angle of 45 degrees. The apparent edges at different times are shown in panels (a) and (b). The lines from the impact point are color-coded for different angles from the target surface. Panels (a) and (c) show the results of the laboratory experiment. Panels (b) and (d) show the results of the simulation for $n_{imp} = 3 \times 10^6$. The dotted lines in panels (c) and (d) are the best-fitting linear functions to the five data points for each angle.

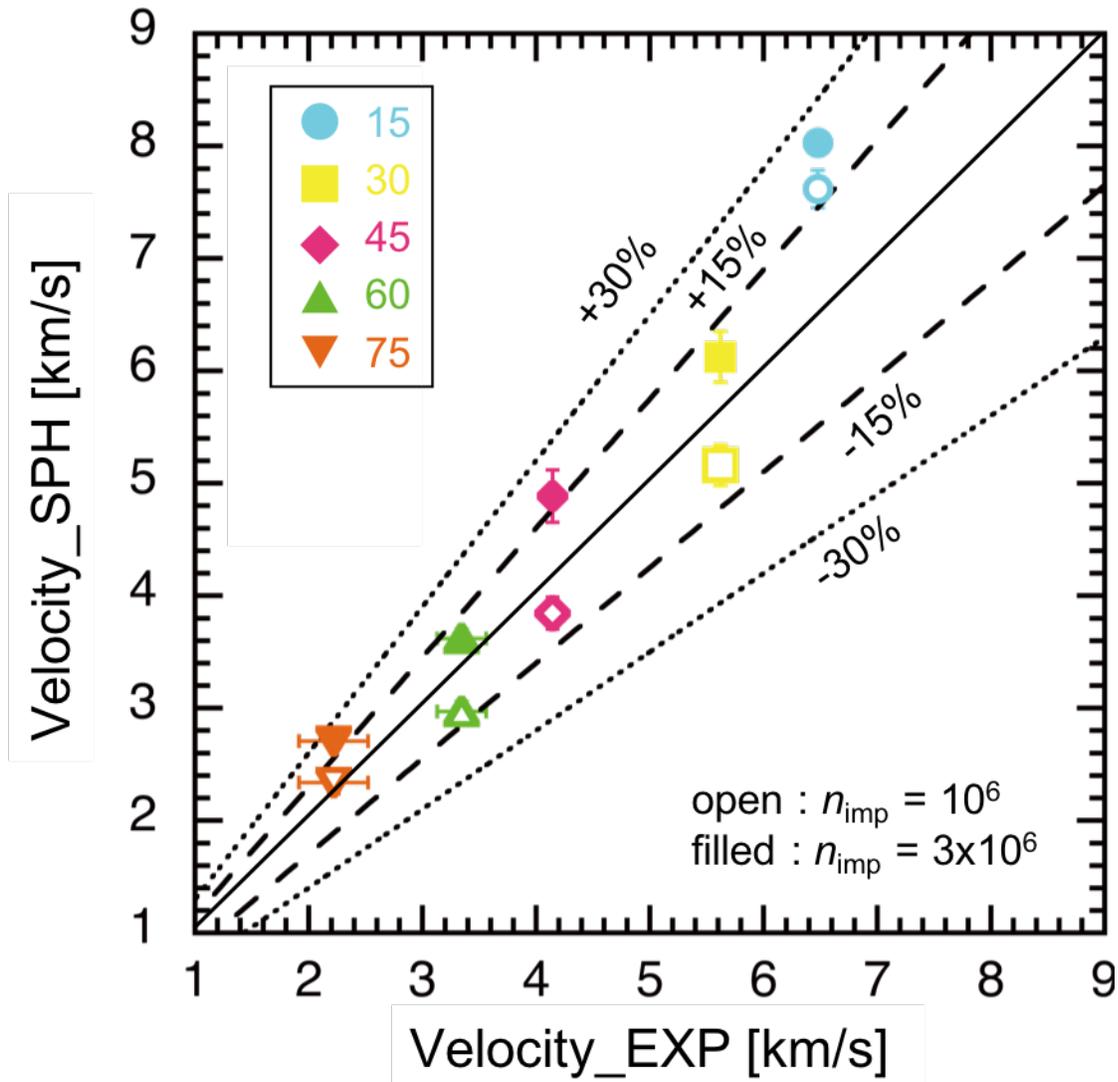

**Figure S7.** Comparison of the moving velocities of the apparent edge between the laboratory experiment and SPH simulation for different angles. This figure shows the results for the shot with an impact velocity of 5.04 km s$^{-1}$ and an impact angle of 45 degrees. The open and filled symbols show the results of the simulations for $n_{imp}=10^6$ and $3\times10^6$, respectively. The numbers next to the symbols represent the angles from the target surface, shown in Figure S6.

**Text S4. Vertical Impact**

S4.1. Results of the Vertical Impact for an Impact Velocity of 4.18 km s$^{-1}$

Figure S8 shows the high-speed video images of the ejected materials for the vertical impact case at impact velocity of 4.18 km s$^{-1}$. An axisymmetric morphology of the ejecta was observed. The rear surface of the projectile was intact until at least $t = 0.6$ μs ($t = 0.52\ t_s$). Figure S9 shows images obtained from the SPH simulations after the data analysis described in Section 2.3 in the main text for different $n_{imp}$ at $t = 2.0$ μs ($t = 1.7\ t_s$), as well as the image of the laboratory experiment. The morphology of the ejecta in the simulation becomes more similar to that in the experiment for increasing $n_{imp}$ for the vertical impact. The image processing described in Text S2 was performed and the apparent edges of the ejecta were extracted. Figure S10(a) and S10(b) show examples of the edges at $t = 1.4, 2.1$, and $2.8\ t_s$ ($t = 1.6$, 2.4, and 3.2 μs, respectively). The morphology of the apparent edge of the ejecta was axially symmetric. The images here thus only show the left-hand side of the edges (i.e., the region at $X < 0$). Figure S10(c) and S10(d) show the distance between the impact point and apparent edges for different angles as a function of time. The distance is almost the same for both results, although the position of the edge in the laboratory experiment tends to be slightly more distant from the impact point than that in the SPH simulation. The difference of the loci of the edge in the SPH simulation and that in the laboratory experiment would be much less if an SPH simulation were conducted at higher spatial resolution. The moving velocities of the edges for different angles were calculated in the same way as for the oblique impact and are shown in Figure S11. The results show that the velocities in the simulation for $n_{imp} \geq 10^6$ are in agreement to within ±30% with the velocities in the laboratory experiment. Especially for $n_{imp} = 3\times10^6$, most velocities are in good agreement, to within ±15% (see the detailed description of the results in Section 3.2 in the main text).

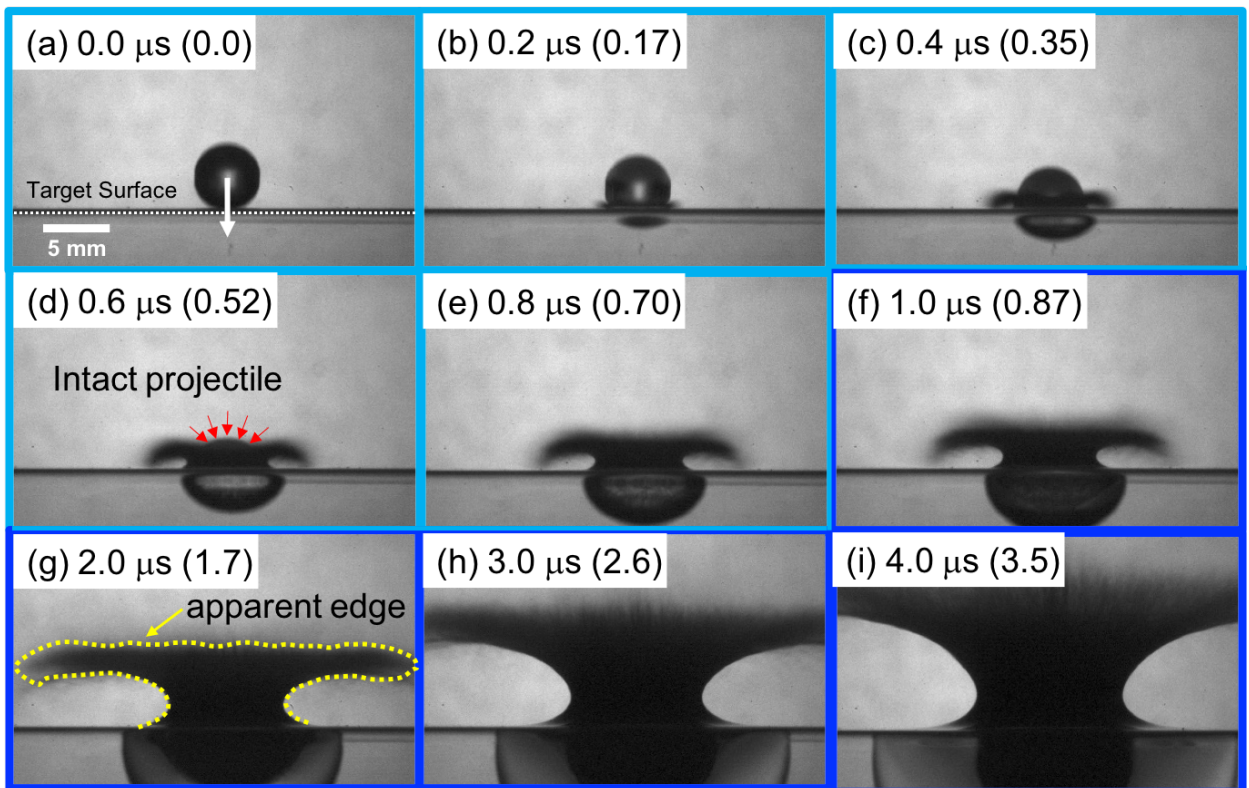

**Figure S8.** High-speed images of the vertical impact for an impact velocity of 4.18 km s$^{-1}$. The timing after initial contact is indicated in each panel. Note that the time intervals between images in panels (a)–(f) and in panels (f)–(i) are different. Numbers in parentheses indicate the scaled time, $t/t_s$, (i.e., the ratio of the real time, $t$, to the characteristic time for projectile penetration, $t_s$). The dotted yellow line in panel (g) shows an example of the apparent edge of the ejecta (see Section 3.2 in the main text).

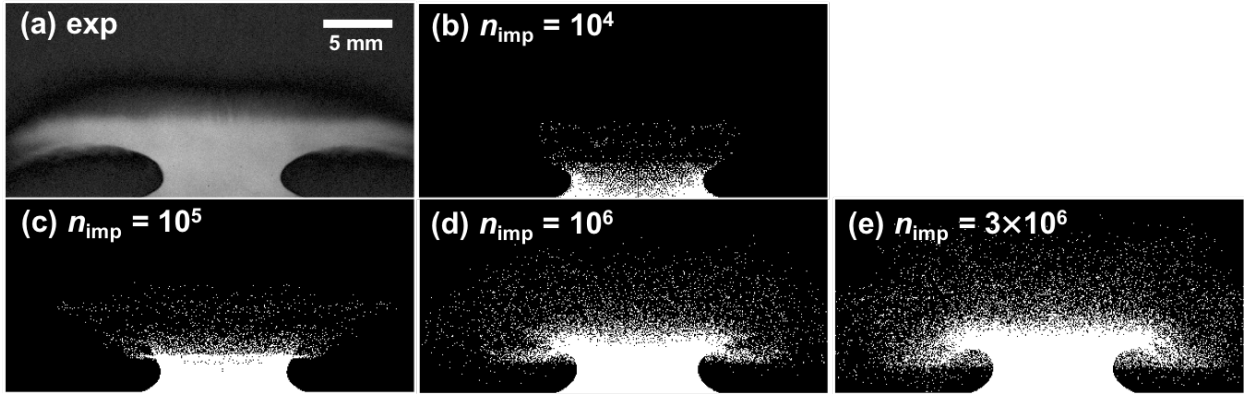

**Figure S9.** Processed images of the SPH simulations in panels (b)–(e) for an impact velocity of 4.18 km s$^{-1}$ for the vertical impact and different $n_{imp}$, as well as the subtracted image of the impact experiment, shown in panel (a). The bottom of each panel corresponds to the target surface. The time shown is 2.0 μs (1.7 $t_s$) after the impact. The image of the impact experiment shown here was obtained from subtraction of the background from the raw image. The horizontal bar in panel (a) indicates the length scale.

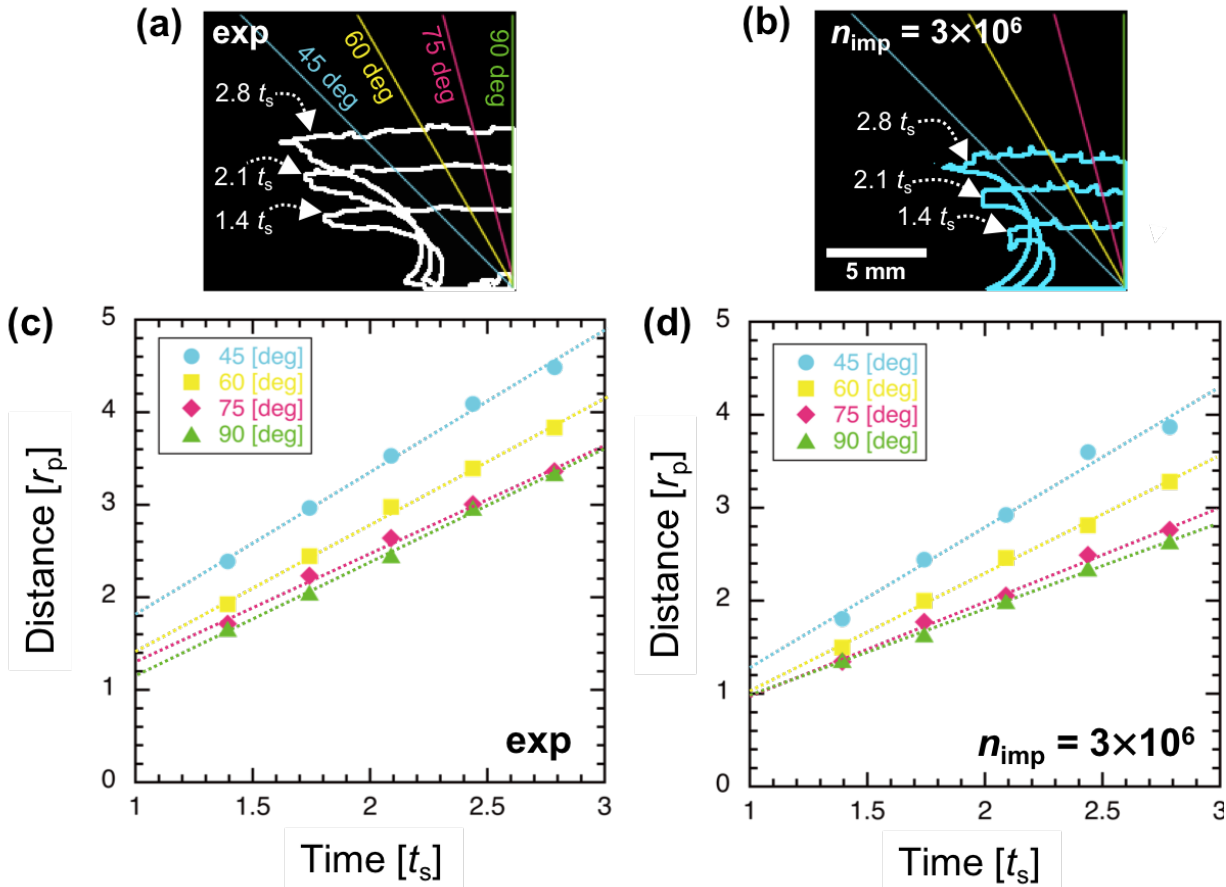

**Figure S10.** Distance between the impact point and apparent edges for different angles as a function of time. This figure shows the results for the shot with an impact velocity of 4.18 km s$^{-1}$ and impact angle of 90 degrees. The apparent edges at different times are represented in panels (a) and (b). Lines from the impact point are color-coded for different angles from the target surface. Panels (a) and (c) show the results of the laboratory experiment. Panels (b) and (d) show the results of the simulation for $n_{imp}=$

$3\times10^6$. The dotted lines in panels (c) and (d) are the best-fitting linear functions to the five data points for each angle.

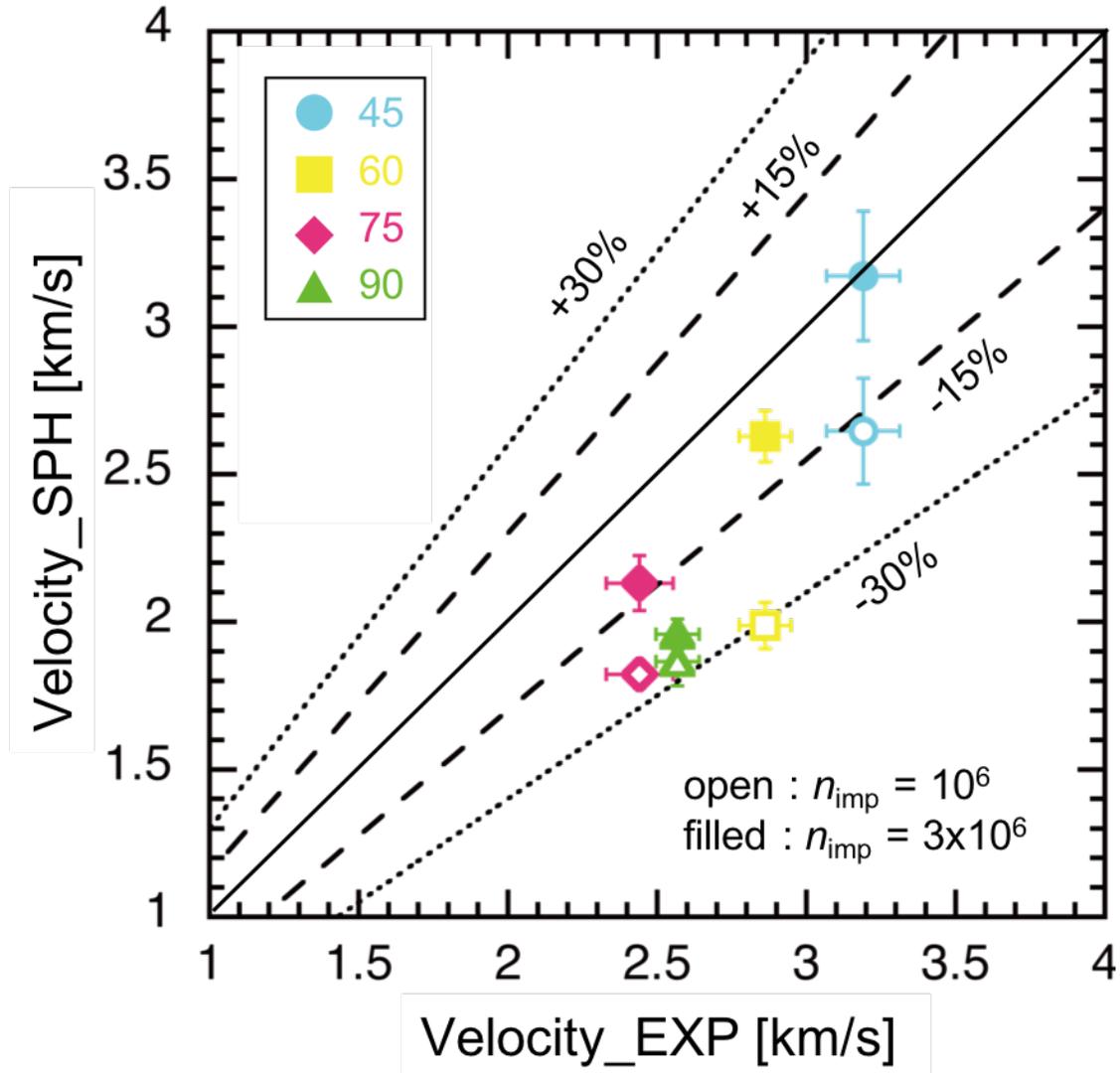

**Figure S11.** Comparison of the moving velocities of the apparent edge between the laboratory experiment and SPH simulation for different angles. This figure shows the results of the shot with an impact velocity of 4.18 km s$^{-1}$ and impact angle of 90 degrees. The open and filled symbols show the results of the simulations for $n_{imp} = 10^6$ and $3\times10^6$, respectively. The numbers next to the symbols represent the angles from the target surface shown in Figure S10.

S4.2. Velocity Distributions for the Vertical Impact

We extracted the SPH particles located on the top edge of the ejecta at $t = 2.6$ μs ($t = 2.3\ t_s$) within 7.5 mm (= the end of the edge in $X$ direction) horizontally from the $Z$ axis because of the axial symmetry of the ejecta. The time used is the same characteristic time as for the oblique impact shown in Figure 11 in the main text. Figure S12 shows the extracted SPH particles, where the color map of the ejection velocity, $v_{ej}$, is normalized by the impact velocity. Ejecta from the target and projectile materials are shown in Figure S12(a) and S12(b) and in Figure S12(c) and S12(d), respectively. Materials from the projectile were distributed more closely to the trajectory of the projectile than materials from the target. Note that the ejection velocities of the particles were lower than the impact velocity, although some SPH particles located on the outside of the apparent edge (> 7.5 mm horizontally from the $Z$ axis) which are not displayed in Figure S12 have velocities higher than the impact velocity.

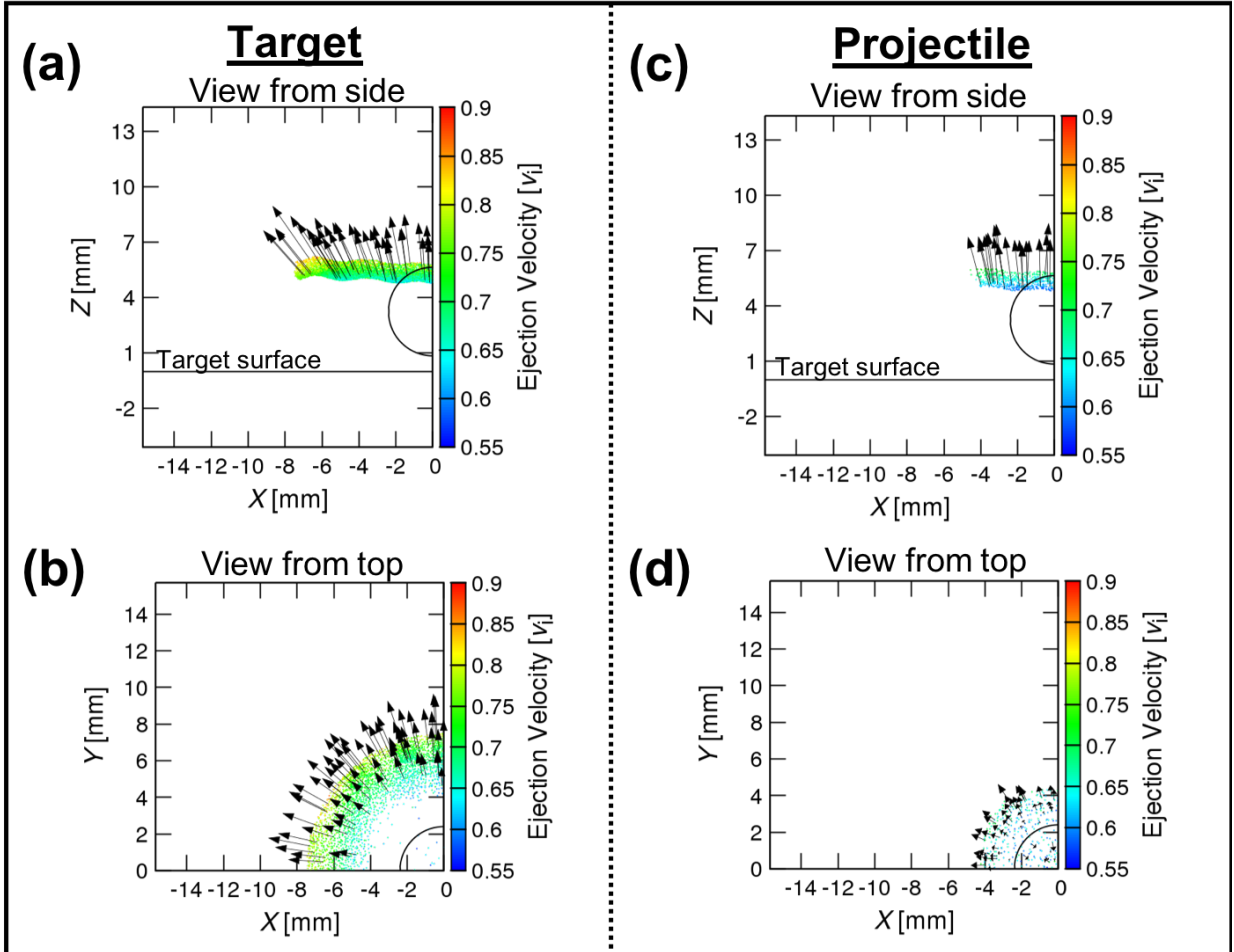

**Figure S12.** Spatial distributions of the extracted SPH particles at $t = 2.6$ μs ($t = 2.3\ t_s$) for an impact velocity of 4.18 km s$^{-1}$ and impact angle of 90 degrees. The SPH particles are color-coded according to their ratios of the ejection velocities to the impact velocity. (a) Target materials in the $X$–$Z$ plane; (b) target materials in the $X$–$Y$ plane; (c) projectile materials in the $X$–$Z$ plane; (d) projectile materials in the $X$–$Y$ plane. The arrows show velocity vectors of the ejection velocities of the particles. The initial positions of the projectiles are shown by the circles. Note that only particles on the apparent edge within 7.5 mm (= the end of the edge in $X$ direction) horizontally from the $Z$ axis have been extracted.

Figure S13 shows the distribution of the direction vectors of the ejection velocities as a function of elevation angle, $\theta$, and azimuth angle, $\phi$. The definitions of $\theta$ and $\phi$ are given in Section 4.1 in the main text. Note that only the ranges $0° \leq \theta \leq 90°$ and $0° \leq \phi \leq 90°$ are shown in Figure S13, because the distribution was an axially symmetric pattern with respect to the $Z$ axis. The ejecta distribution in this diagram is characterized as follows. Most of the target materials were distributed across $45° \leq \theta \leq 75°$.

The ejection velocity decreases with increasing $\theta$. In contrast, projectile materials were mostly distributed at high angles, $70° \leq \theta \leq 90°$, and some proceed in the direction of the projectile's trajectory (i.e., the $+Z$ axis).

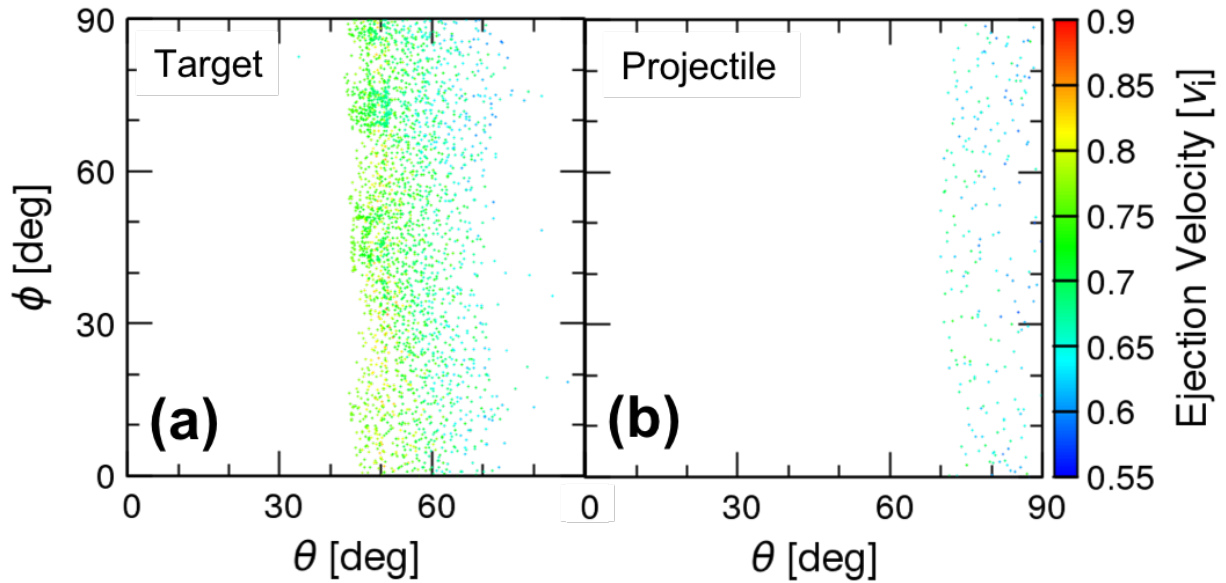

**Figure S13.** Ejection velocity distributions of the extracted SPH particles at $t = 2.6$ μs ($t = 2.3\ t_s$) shown in Figure S12 as a function of $\phi$ and $\theta$. The definitions of the elevation, $\theta$, and azimuth, $\phi$, are illustrated in panel (a) of Figure 12 in the main text. The SPH particles from the target materials [panel (a)] and from the projectile [panel (b)] are color-coded depending on their velocity ratios of the ejection velocities to the impact velocity, as in Figure S12.

Initial loci of the extracted SPH particles from the target in Figure S12 are shown in Figure S14. Note that some of the extracted particles did not continuously spread out possibly because they were not extracted from a perfectly smooth apparent edge due to the image processing. The particles were launched from approximately a projectile radius away from the impact point. The velocities of the particles from the target tend to decrease with increasing depth of the initial positions. These particles are located in the first to sixth layers of the target, which implies that as for the oblique impact our SPH model for vertical impact can also reproduce the hydrodynamic behavior of the particles initially located even near the surface (see the detailed description in Section 4.1 in the main text).

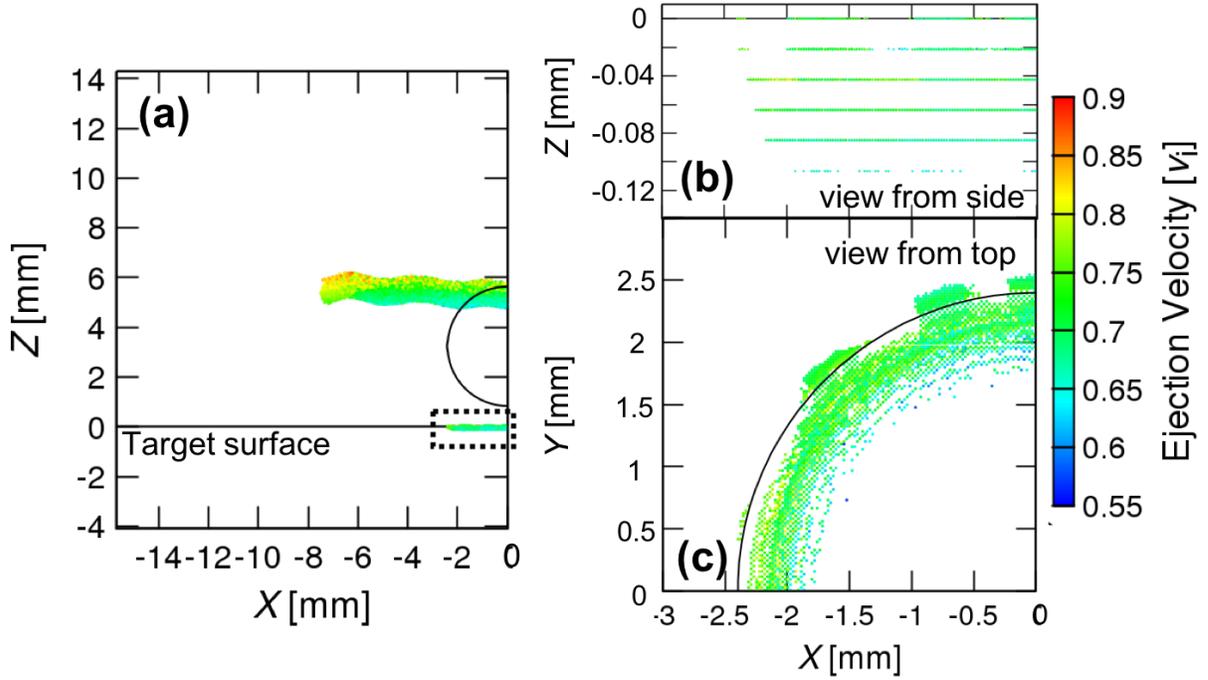

**Figure S14.** Initial positions of the extracted SPH particles in the target materials shown in Figure S12. SPH particles are color-coded according to their ratios of the ejection velocities to the impact velocity. (a) Initial positions of the extracted SPH particles in the $X$–$Z$ plane at $t = 2.6$ μs ($t = 2.3\ t_s$). The initial positions of the projectiles are shown as circles. Enlarged views of the area surrounded by the dotted rectangle are shown in panels (b) and (c). Panels (b) and (c) show the initial positions of the extracted SPH particles in the $X$–$Z$ and $X$–$Y$ planes, respectively. The solid line in panel (c) indicates the initial position of the projectile.

**Text S5. Ejection velocity and mass distribution**

Figure S15 shows the cumulative mass of ejecta with velocities greater than a given ejection velocity, $v_{ej}$, as a function of ejection velocity for an impact velocity of 3.56 km s$^{-1}$ in the oblique impact simulation with $n_{imp} = 3 \times 10^6$. Note that this figure was obtained from all ejecta moving above the target surface at the end of the calculation (i.e., 4.8 μs after impact). Note also that the number of particles from the target characterized by ejection velocities greater than twice the impact velocity is more than 2000. Thus, most particles ejected at such high velocities are not expected to be artifacts.

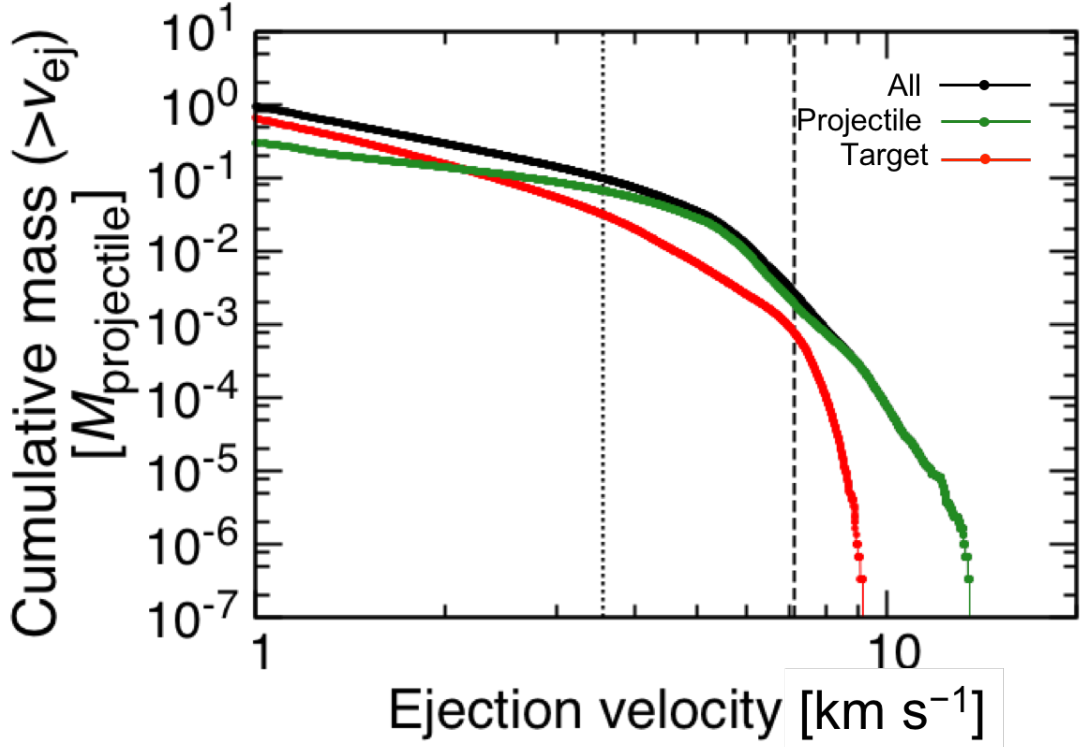

**Figure S15.** Cumulative ejecta mass normalized by projectile mass in the oblique impact simulation. The cumulative masses of the target and projectile materials, as well as of both materials, are shown separately as red, green, and black lines, respectively. The dotted and dashed vertical lines indicate impact velocities, $v_i = 3.56$ km s$^{-1}$ and $2\,v_i$.

**Text S6. Accuracy of the EOS Model in the Simulations**

It is widely recognized that the accuracy of the Tillotson EOS in the expansion region ($\rho/\rho_0 \ll 1$ and $E > E_{iv}$, where $\rho$, $\rho_0$, $E$, and $E_{iv}$ are the density, the reference density = 1200 kg m$^{-3}$, the internal energy, and the internal energy at the incipient vaporization, respectively) is relatively low with respect to other EOS models, such as ANEOS (Thompson and Lauson, 1972). This is because of the simplified treatment of the phase change from the condensed phase to the gas phase. We must address the possibility that the observed ejecta with extremely high ejection velocities originate from numerical artifacts owing to uncertainties in the Tillotson EOS. Figure S16 shows the time variations of the density and particle velocity of typical SPH particles with ejection velocities greater than the impact velocity. We confirmed that the density at the time when the particle reaches its maximum velocity is close to the reference density (i.e., the solid density) in the simulation. We did not observe any significant acceleration after decompression ($\rho/\rho_0 < 1$) because of a much lower pressure (weaker pressure gradient to vacuum) than at the time when the acceleration occurs. Although the density of the ejected SPH particles decreases numerically with time in the SPH simulations after transition to a linear uniform motion, this density decrease would occur in any SPH code, regardless of the choice of EOS model. This is because the interparticle distance between SPH particles gradually increases with time after ejection, resulting in the numerically low "spatial" density, defined as the mass of each SPH particle divided by the volume it occupies. Most of the acceleration, however, occurs in the compressed region of the Tillotson EOS ($\rho/\rho_0 > 1$) in our simulations. It is well known that the accuracy of the Tillotson EOS in the compressed region is relatively high because the Tillotson parameters are determined based on the physical properties of the solid and shock Hugoniot parameters. Consequently, we can rule out the possibility that the high ejection velocities observed in the simulation are artifacts caused by the simplified treatment of the vaporization in the Tillotson EOS.

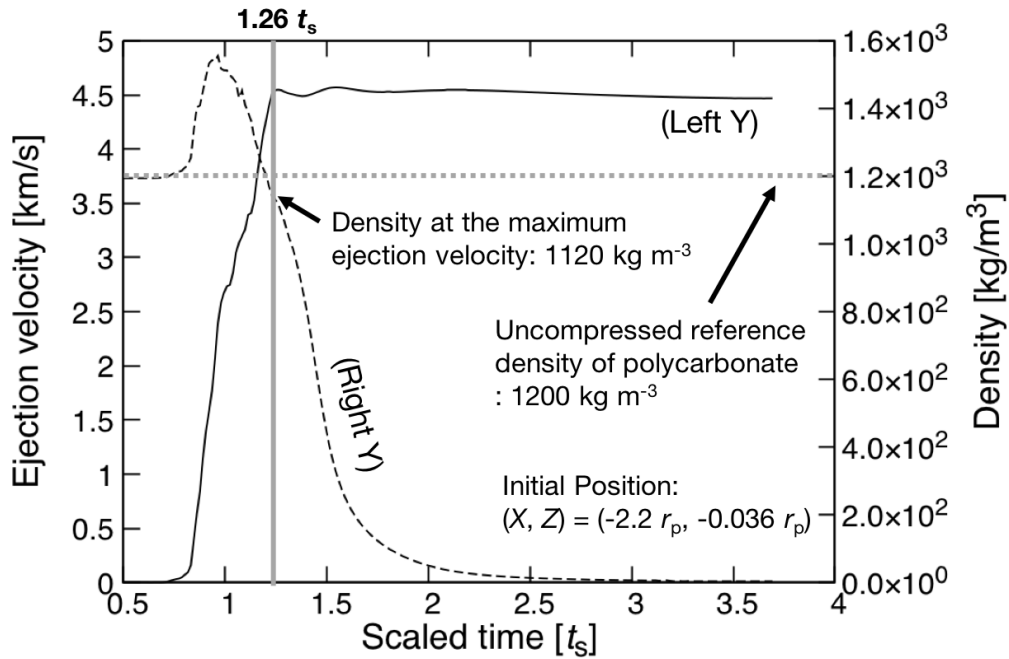

**Figure S16.** Temporal variations of the ejection velocity (solid line; left *Y* axis), density and pressure (dashed line; right *Y* axis) for typical SPH particles with ejection velocities greater than the impact velocity. The gray horizontal dotted and vertical solid lines show the reference density of uncompressed polycarbonate (= 1200 kg m$^{-3}$), and the time when the particle reached its maximum ejection velocity (= 1.26 $t_s$). The values in parentheses indicate the initial (*X*, *Z*) position of the particles in units of the projectile radius, $r_p$.